\documentclass[twocolumn,astrosymb]{aastex631}

\usepackage{textcomp}
\usepackage{amssymb}
\usepackage{amsmath}
\usepackage{comment}
\usepackage{wasysym}
\usepackage{CJK}
\usepackage{verbatim}
\usepackage{enumerate}

\shorttitle{Spatially resolved outflows in NGC 7469}
\shortauthors{Xu \& Wang}

\graphicspath{{./}}

\begin{document}
\begin{CJK*}{UTF8}{gbsn}

\title{Spatially Resolved Ionized Outflows Extending to $\sim$2 kpc in Seyfert 1 Galaxy NGC 7469 Revealed by VLT/MUSE}

\correspondingauthor{Junfeng Wang}
\email{jfwang@xmu.edu.cn}

\author[0000-0003-0970-535X]{Xiaoyu Xu (许啸宇)}
\affil{Department of Astronomy, Xiamen University, Xiamen, Fujian 361005, China}

\author[0000-0003-4874-0369]{Junfeng Wang}
\affiliation{Department of Astronomy, Xiamen University, Xiamen, Fujian 361005, China}

\begin{abstract}

The Seyfert 1 galaxy NGC 7469 possesses a prominent nuclear starburst ring and a luminous active galactic nucleus (AGN). 
Evidence of an outflow in the innermost nuclear region has been found in previous works.  
We detect the ionized gas outflow on a larger scale in the galaxy using the archival VLT/MUSE and {\em Chandra} observations. 
The optical emission lines are modeled using two Gaussian components, and a non-parametric approach is applied to measure the kinematics of [O~{\sc{iii}}] and $\rm H\alpha$ emitting gas. 
Line ratio diagnostics and spatially resolved maps are derived to examine the origin of the outflow.  
The kpc-scale kinematics of [O~{\sc{iii}}] is dominated by a blueshifted component whereas velocity map of $\rm H\alpha$ shows a rotational disk with complex non-rotational substructure.  
The starburst wind around the circumnuclear ring is confirmed, and we find evidence of an AGN-driven outflow extending to a radial distance of $\rm \sim2$ kpc from the nucleus, with a morphology consistent with a nearly face-on ionization cone. 
The previously reported circumnuclear outflow resembles part of the bright base.  
We derive mass and energy outflow rates for both the starburst wind and the AGN-driven outflow. 
The estimated kinetic coupling efficiency of the kpc-scale AGN outflow is $\dot{E}_{\rm out}/L_{\rm bol}\sim 0.1\%$, lower than the threshold predicted by the ``two-stage'' theoretical model for effective feedback. 
Our results reinforce the importance of spatially resolved study to disentangle feedback where AGN and starburst coexist, which may be common during the cosmic noon of black hole and galaxy growth.

\end{abstract}

\keywords{Seyfert galaxies (1447) --- Galaxy winds (626) --- Interstellar medium (847) --- Luminous infrared galaxies (946)}

\section{Introduction} \label{sec:intro}

Galactic outflows are predicted by theoretical models of galaxy evolution and can be found in a variety of galaxies at different redshifts. 
They can be driven either by starburst processes or by an active galactic nucleus. 
It is generally expected that these outflows can have a profound impact on the environments and evolution of host galaxies by interacting with the interstellar and intergalactic medium (ISM and IGM, respectively). 
In negative feedback scenario, outflows can heat and expel cold gas so that the star formation (SF) will be suppressed by short of gas supplies \citep[e.g.][]{1998A&A...331L...1S,2008ApJS..175..356H,2012ARA&A..50..455F,2012RAA....12..917S}. 
Alternatively some works have suggested that outflows can also trigger or enhance SF processes, known as positive feedback \citep[e.g.][]{2013ApJ...772..112S,2017MNRAS.468.4956Z}.  

Outflows driven by SF generally have higher impact in low-mass galaxies \citep[e.g.][]{2005ARA&A..43..769V}, which can prevent dwarf galaxies from forming and growing \citep[e.g.][]{1998A&A...331L...1S,2012RAA....12..917S}. 
In nearby galaxies, especially in luminous and ultra-luminous infrared galaxies (LIRGs and ULIRGs, respectively), and starbursts galaxies, these outflows have been identified \citep[e.g.][]{2000ApJS..129..493H,2005ApJS..160..115R}. On the other hand, outflows powered by AGN, which can take the forms of highly collimated jets or winds with larger opening angles, are believed to be the main conveyor of the AGN feedback. Powerful jets produced by massive radio galaxies can heat and maintain the intracluster medium (ICM) in the hot phase, which has been observed in several galaxy clusters \citep[e.g.][]{2012ARA&A..50..455F}. 
Strong AGN-driven outflows are usually found in luminous AGNs, where the supermassive black hole (SMBH) in the center of galaxies is in a high accretion state \citep[e.g.][]{2015Natur.519..436T,2021MNRAS.504.3890D}. 

Although unambiguous evidence of outflows driven by SF or AGN has been found in observations, the exact physical processes hence the ultimate impact on their host galaxies remain an open question. Spatially resolved and multi-wavelength studies are becoming powerful in solving this issue over the years \citep[e.g.,][]{2017ApJ...844...69M,2017MNRAS.467.2612W,2018A&A...619A..74V,2019A&A...622A.146M}. 
In the optical and the X-ray band, extended narrow emission line regions (ENLRs) in nearby Seyfert galaxies have been demonstrated to be a useful probe of outflows, jet-ISM collision, starburst, and/or AGN-photoionization \citep[e.g.,][]{2009ApJ...704.1195W,2012ApJ...756..180W,2014ApJ...781...55W,2018ApJ...855..131F}. Warm ionized gas outflows ($\rm T\eqsim 10^{4}K$) traced by [O~{\sc{iii}}] and [N~{\sc{ii}}]+H$\alpha$ are prevalent in AGNs and star-forming galaxies. Integral field spectroscopy (IFS) provides more detailed information such as spatial resolved kinematics, ionization structure and morphology of ionized outflows, which enables better understanding of physical processes in outflows \citep[e.g.][]{2014MNRAS.441.3306H,2017MNRAS.467.2612W,2018A&A...619A..74V,2020AJ....159..167L,2021MNRAS.504.3890D,2021ApJ...913..139G}.

NGC 7469 is a nearby \citep[$z=0.01632$,][]{1996ApJS..106...27K} Seyfert 1.5 \citep[][]{2006A&A...455..773V} galaxy with the SMBH mass $\rm 1.1\times 10^{7}\ M_{\astrosun}$, measured by the latest reverberation mapping \citep[][]{2021ApJ...918...50L}. 
This galaxy is classified as LIRG due to its high IR luminosity \citep[$L_{\rm 8-1000\ \mu m}=10^{11.6}\ L_{\rm \astrosun}$,][]{2003AJ....126.1607S}.

There exists a well-known circumnuclear starburst ring in NGC 7469 detected in almost all wavelength \citep[e.g.][and references therein]{2020ApJ...898...75I}, likely triggered by interaction with the companion IC 5283 galaxy ($\sim1.\arcmin3$ away from NGC 7469).  The outer and inner radius of this starburst ring are $\rm \sim 2.\arcsec5$ and $\rm \sim0.\arcsec7$, respectively \citep{1995ApJ...444..129G}. 
Most star clusters in the ring are relatively young \citep[$\rm <30\ Myr$,][]{2007ApJ...661..149D}. A circumnuclear disk (CND) with a radius of $\rm \sim 150\ pc$ \citep[e.g.][]{2004ApJ...602..148D,2015ApJ...811...39I} is presented in the center of NGC 7469, previously studied as an X-ray dominated region produced by AGN \citep{2020ApJ...898...75I}. 
\citet{2004ApJ...602..148D} detected a bar or a pair of loosely wound spiral arms between the starburst ring and the CND based on the IRAM CO(2-1) data.  \cite{2015ApJ...811...39I} further suggested that these structures are possibly several spiral features according to lacking corresponding stellar structure based on their ALMA observations.
In the radio, an unresolved compact source associated with the AGN and a ring emission originated from the starburst ring have been found by \cite{1991ApJ...381...79W} using Very Large Array (VLA).
Further radio observations with high spatial resolutions have revealed a core-jet structure in the nuclear region on the sub-arcsecond scale\citep[][]{2003ApJ...592..804L,2006ApJ...638..938A,2010MNRAS.401.2599O}. Such small scale radio jet is commonly seen in other nearby radio quiet Seyfert galaxies \citep[e.g.,][]{1993MNRAS.263..471P}. No evidence of jet-ISM interaction was found for NGC 7469 in the literature.

In the X-ray and ultraviolet (UV), spatially unresolved AGN winds have been ubiquitously detected in absorption lines of highly ionized gas known as warm absorbers \citep[e.g.,][]{1994MNRAS.268..405N}. 
With 160 ks {\it XMM}-Newton Reflection Grating Spectrometers (RGS) spectrum, \citet{2007A&A...466..107B} reported AGN winds with velocities of -580 to -2300 $\rm km\ s^{-1}$, which is consistent with previous UV observation \citep{2005ApJ...634..193S}. 
Later \cite{2017A&A...601A..17B} confirmed these results with 640 ks RGS spectrum with velocities of -650 to -2050 $\rm km\ s^{-1}$, and they also found a collisionally ionized component at $\rm kT=0.35\ keV$ that they associated with the nuclear starburst ring since its luminosity was expected from far-infrared SF emission. 
\cite{2018A&A...615A..72M} also reported AGN winds with velocities ranging from -400 to -1800 $\rm km\ s^{-1}$ using the {\em Chandra} HETGS. 
They also discovered diffuse soft (0.2--1 keV) X-ray emission extending up to $12\arcsec$, which is consistent with coronal emission from nuclear starburst ring in NGC 7469.

In the last century, \cite{1986ApJ...310..121W} have reported that [O~{\sc{iii}}] emission lines in the circumnuclear region show strong, blueward-slanting profile asymmetries using long-slit mapping, which present non-rotational kinematics. 
Using the VLT/SINFONI IFS, \cite{2011ApJ...739...69M} detected a biconical outflow with a maximum velocity of $\rm \sim130\ km\ s^{-1}$ and extending up to 380 ± 25 pc from the AGN, in the infrared coronal line [Si VI]$\rm \lambda 1.96\ \mu m$. 
\cite{2020MNRAS.493.3656C} did not report the biconical outflow in [N~{\sc{ii}}]+H$\alpha$ using GTC/MEGARA but presented a non-rotational turbulent component possibly associated with an outflow. 
Recently, \cite{2021ApJ...906L...6R} have found evidence of two outflow components in the VLT/MUSE data. 
The slower component extends across most of the nuclear starburst ring with a median velocity of -284 $\rm km\ s^{-1}$, which is possibly driven by massive starforming. 
The faster component with a median velocity of -581 $\rm km\ s^{-1}$, in the western region between the AGN and the starburst ring, is likely AGN-driven. Their work focused on only the bright innermost nuclear region of NGC 7469, whereas ionized gas on the larger scale (with radius $\rm r>2.\arcsec5$) was unexplored possibly due to the lower signal-to-noise ratio (S/N) in those spaxels. 

In this paper, we report galactic outflow features in NGC 7469 using the same Multi-Unit Spectroscopic Explorer (MUSE) data as \cite{2021ApJ...906L...6R}, utilizing adaptive binning methods to improve the S/N. 
In Section \ref{sec:2}, we describe the methodology including observations, data reduction, and the approach used to characterize emission lines. 
Main results including kinematics maps, ionization and excitation diagnostic diagrams and maps, and estimations of outflow rates are described in Section \ref{sec:3}. 
Discussions and brief conclusions are presented in Section \ref{sec:4} and Section \ref{sec:5}, respectively.

\section{Observations and Data Analysis}
\label{sec:2}
\subsection{MUSE data}
\label{sec:2.1}

The MUSE \citep[][]{2010SPIE.7735E..08B} is an IFS located at the VLT of the European Southern Observatory (ESO). In the wide field mode (WFM), MUSE provides a large field of view (FoV) of nearly 1 arcmin$^{2}$ with spatial sampling of $\rm 0.\arcsec2$ pixel$^{-1}$, and a wavelength coverage of $\rm 4650-9300\, \AA$ with a mean resolution of R $\sim$ 3000. NGC 7469 was observed in WFM-NOAO mode on August 19, 2014 during the science verification run with an exposure time of 2.4 ks and mean seeing of $\rm 1.\arcsec23$ \citep[see also][]{2016MNRAS.455.4087G,2021ApJ...906L...6R}. 

We downloaded the processed data from ESO Archive Science Portal \footnote{\url{http://archive.eso.org/scienceportal/home}} and fitted the datacube using the LZIFU pipeline \citep[][]{2016Ap&SS.361..280H}, which is an IDL toolkit including the Penalized Pixel-Fitting method \citep[PPXF;][]{2004PASP..116..138C} and the MPFIT \citep[][]{2009ASPC..411..251M}. 
To cover the main emission lines in optical band, for example, H$\beta$, [O~{\sc{iii}}]$\lambda\lambda$4959,5007, H$\alpha$, [N~{\sc{ii}}]$\lambda\lambda$6548,6583, and [S~{\sc{ii}}]$\lambda\lambda$6716,6731, a fitting range of $\rm 4770-7550\,\AA$ was set. 
The redshift $z=0.01632$ measured by \cite{1996ApJS..106...27K} is used to correct the observed wavelength to the rest frame.
Spaxels with mean S/N $\leq$1 (calculated from the spectrum between 5000 to 7000 $\rm \AA$) were excluded. 
We rebinned the remaining spaxels using the Voronoi binning method \citep[][]{2003MNRAS.342..345C} to achieve a S/N threshold of 25. 
 
\begin{figure*}[htbp!]
	\includegraphics[width=\textwidth]{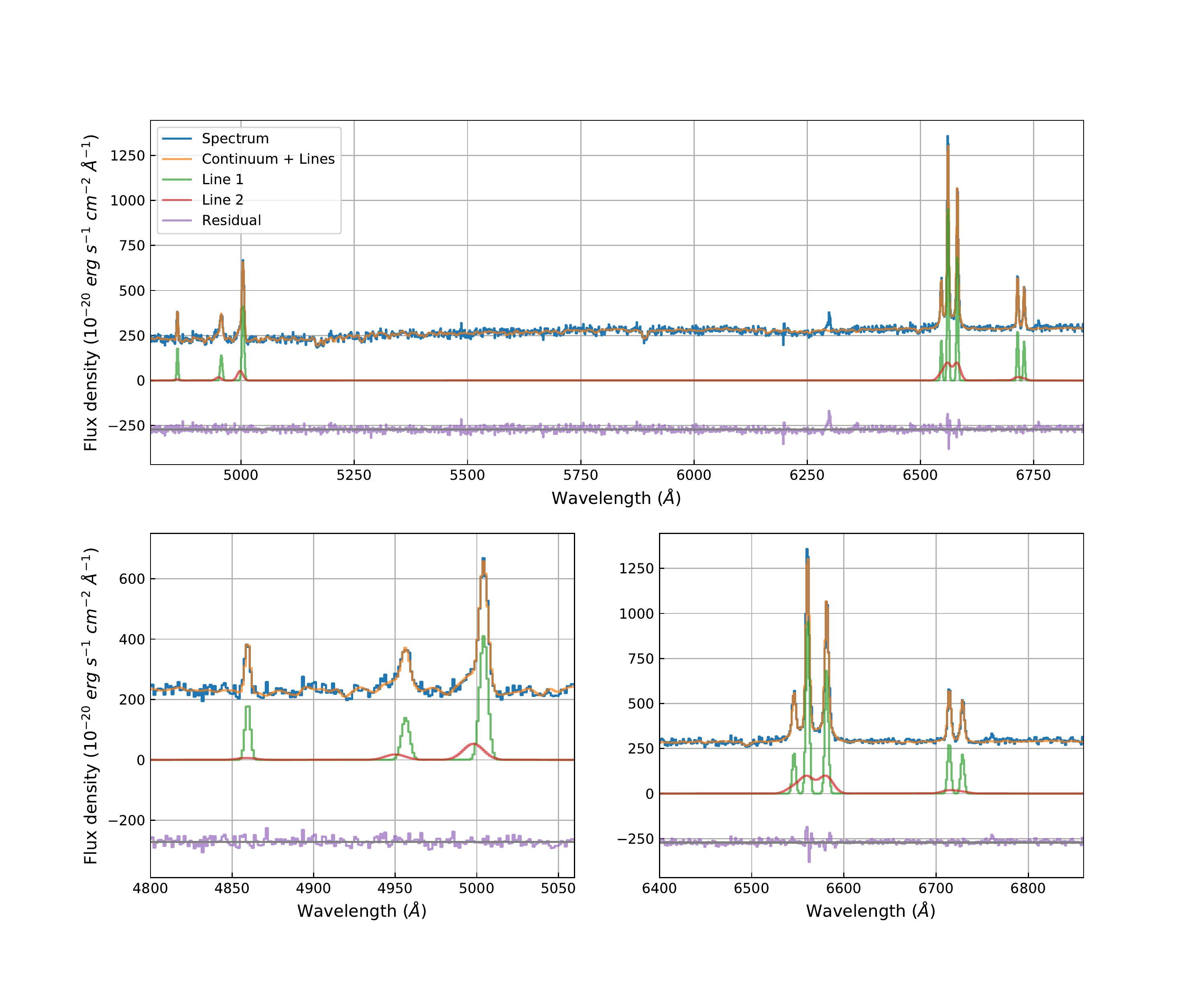}
    \caption{An example of spectra (shown in blue) with the best fit model (continuum and emission lines shown in orange) from a spaxel at $\Delta$RA $=\rm -3.\arcsec6$ and $\Delta$Dec $=\rm -0.\arcsec2$ from the nucleus. 
    The lower left panel shows the spectrum zoomed in on the emission lines H$\beta$, [O~{\sc{iii}}], and the lower right panel H$\alpha$, [N~{\sc{ii}}], and [S~{\sc{ii}}]. Both model components (line 1 and 2) of each emission line are shown in green and red, respectively. For better visualization, the residual (shown in purple) is arbitrarily shifted below zero.}
    \label{fig:spec}
\end{figure*}

The [O~{\sc{iii}}] emission lines show apparent blue asymmetry in line profile, which is indicative of ionized gas outflow.
The Balmer emission lines also present asymmetric profiles but weaker than the [O~{\sc{iii}}], which is more likely dominated by rotation of the gas disk \citep[also noted by][]{2021ApJ...906L...6R}.
Such features in kinematics are common in nearby AGN hosts \citep[e.g.][]{2016ApJ...819..148K}.
When fitting H$\beta$, [O~{\sc{iii}}], H$\alpha$, and [N~{\sc{ii}}], two Gaussian components were used in the whole FoV for asymmetric line profiles.
Note that the [O~{\sc{iii}}] doublets were fitted independently, and when fitting H$\beta$, H$\alpha$, and [N~{\sc{ii}}] emission lines, their center wavelength and velocity dispersion $\sigma$ for the same component share the same value.
Because [N~{\sc{ii}}]$\lambda\lambda$6548,6583 are strongly blended with H$\alpha$, it is reasonable to fix their velocity and $\sigma$ with the same values of Balmer emission lines.
Two components of emission lines were sorted using $\sigma$, i.e., the one with smaller $\sigma$ was component 1 (the narrow component) and the one with larger $\sigma$ was component 2 (the broad component). 
The [S~{\sc{ii}}]$\lambda\lambda$6716,6731 doublets show very weak asymmetric line profile, and only one Gaussian component is used to fit them for estimating the electron density.
Figure~\ref{fig:spec} shows one example of the spectra and models. 
The broad component of [O~{\sc{iii}}] evidently show larger wavelength shifts compare with other lines. 

In this work, we focus on the larger scale ionized gas of NGC 7469 with FoV of $\rm \sim14\arcsec \times \rm 14\arcsec$ than \cite{2021ApJ...906L...6R} with FoV of $\rm \sim5\arcsec \times \rm 5\arcsec$. 
To improve robustness of our results, we rejected components and spaxels with S/N $<$ 5 for all fitted emission lines.

To correct the extinction effect, the Balmer decrement (the flux ratio of H$\alpha$ and H$\beta$) was used. The attenuation law from \cite{1999PASP..111...63F} with $\rm R_{V}=3.12$ and an intrinsic Balmer decrement $\rm H\alpha / H\beta=2.86$ \citep[][]{2006agna.book.....O} were adopted for galactic diffuse ISM.

\subsection{Non-parametric approach to emission line fitting}
\label{sec:nonpar}

\begin{figure}
	\includegraphics[width=\columnwidth]{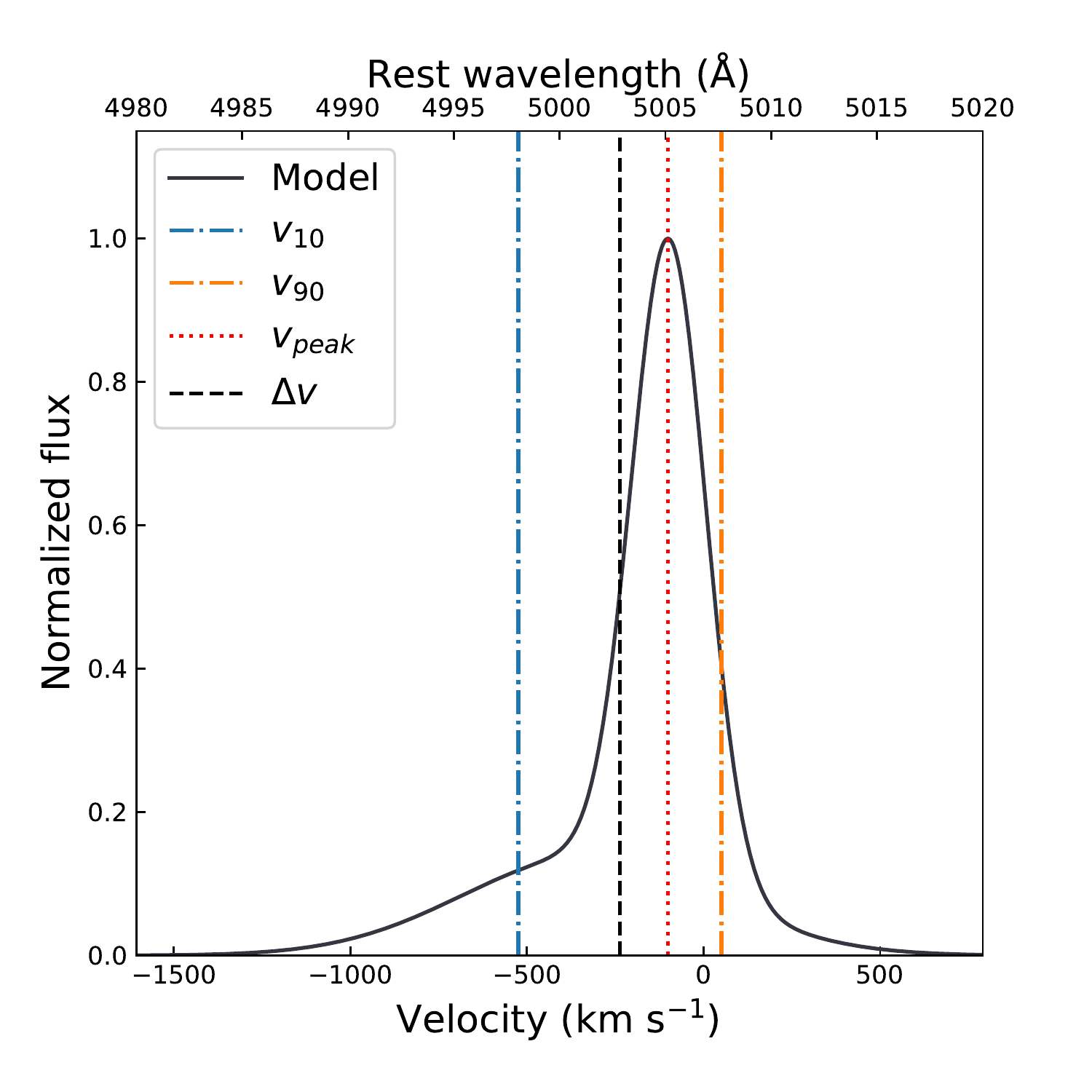}
    \caption{An illustration of the non-parametric method. 
    The solid line is a synthetic model of [O III]$\lambda$5007 extracted from a nuclear spaxel ($\Delta$RA $=\rm 0\arcsec$ and $\Delta$Dec $=\rm -0.\arcsec2$), which is composed of two Gaussian components without the stellar continuum. 
    $v_{\rm 10}$ and $v_{\rm 90}$ denote the velocities at the 10th and 90th percentiles, respectively.
    $\Delta v$ is defined as $(v_{\rm 90}+v_{\rm 10})/2$. $v_{\rm peak}$ is the velocity at the line peak.
    }
    \label{fig:non-par}
\end{figure}

A non-parametric method is applied to characterize the line width and the velocity of emission lines within each spaxel for MUSE data following \cite{2014MNRAS.441.3306H}. For simplicity, the synthetic model with two Gaussian components (fitted by LZIFU) of each emission line was used instead of the data to provide the kinematics. 
As shown in Figure~\ref{fig:non-par}, we use $W_{80}=v_{90}-v_{10}$ and $\Delta v=(v_{\rm 90}+v_{\rm 10})/2$ to represent the line width and the velocity offset of the emission line model, where $v_{\rm 10}$ and $v_{\rm 90}$ are the velocities at the 10th and 90th percentiles.

\subsection{Chandra X-ray data}
\label{sec:xray}

\begin{figure}
	\includegraphics[width=\columnwidth]{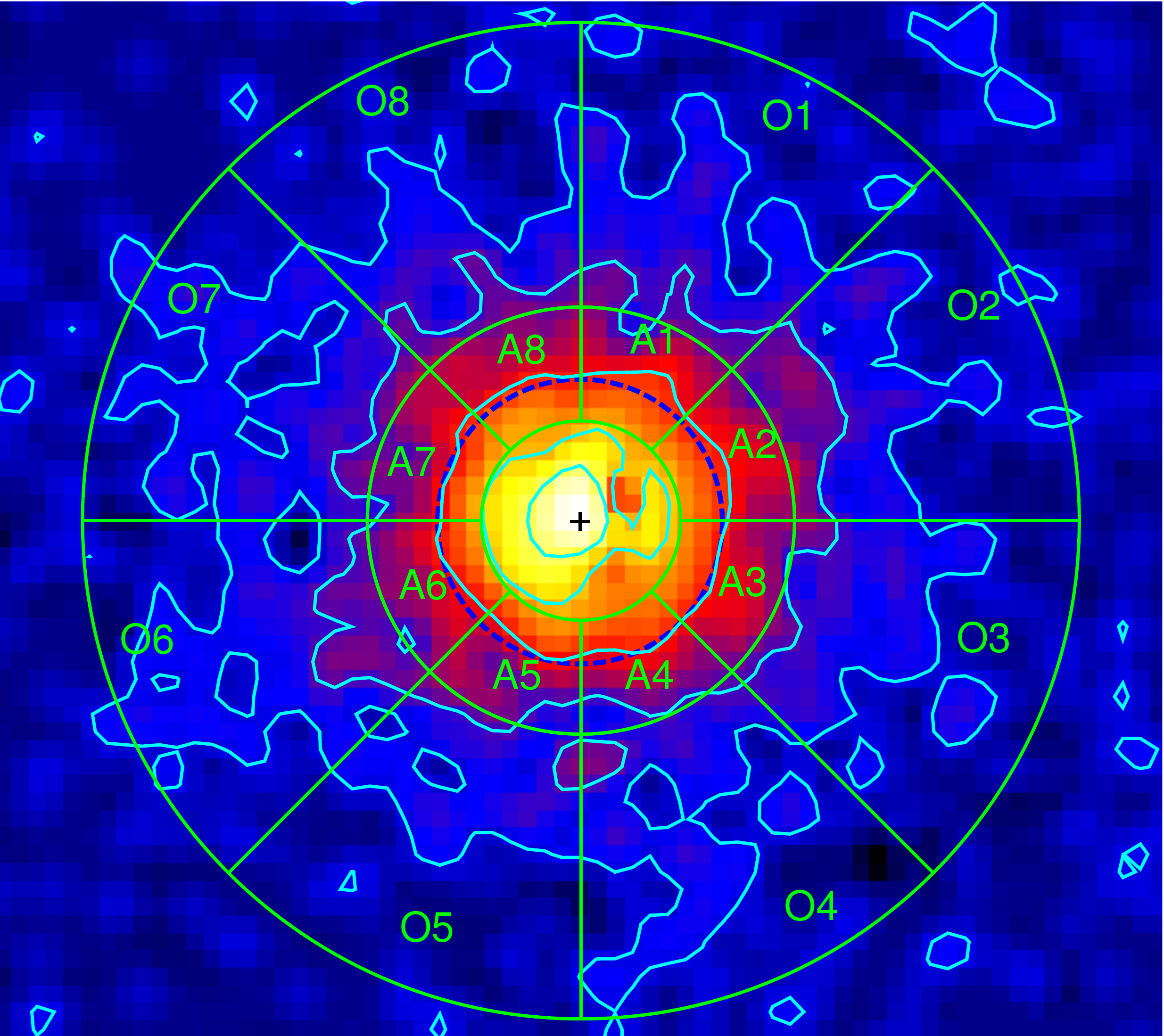}
    \caption{The 0.2--1.0 keV X-ray image with a PSF model of the nucleus subtracted is superposed with corresponding contours. 
    The image is smoothed with the Gaussian function ($\rm \sigma = 1.0$ pixel). 
    Levels of contours are 0.3, 1.0, 5.0, 18, and 50 counts, respectively. 
    The blue dash circle denotes the starburst ring with a radius of $\rm 2\arcsec$ and the black cross marks location of the nucleus.
    The radius of three green circles are $1.\arcsec4$, 3$\arcsec$, and 7$\arcsec$, respectively.
    The two ring regions are separated to 8 sectors and marked as A1--A8 (around the starburst ring, 1.\arcsec4--3$\arcsec$) and O1--O8 (outside the ring, 3$\arcsec$--7$\arcsec$), respectively.
    North is to the top and east is to the left.
    }
    \label{fig:xray}
\end{figure}

NGC 7469 was observed with ACIS-S with HETG onboard {\em Chandra} X-ray Observatory for ten times from 2002 to 2016. Previously diffuse soft X-ray emission in NGC 7469 has been found by \cite{2018A&A...615A..72M} using the {\em Chandra} observation with longest exposure (79ks, Obs. ID:2956). We obtained all available {\em Chandra} observations of NGC 7469 from {\em Chandra} Data Archive\footnote{\url{https://cda.harvard.edu/chaser/}}, reprocessed and merged them using CIAO 4.13 \citep[][]{2006SPIE.6270E..1VF} with CALDB version 4.9.4. The total exposure time is 384 ks. We use this data set to compare the diffuse soft X-ray emission \citep[here denotes 0.2--1.0 keV band, in which the observation brightness profile show excess than PSF of][see their Figure 11]{2018A&A...615A..72M} with the optical IFU data.
Following \cite{2018A&A...615A..72M}, a point spread function (PSF) for the X-ray image is simulated using {\em Chandra} Ray Tracer (ChaRT) \citep[][]{2003ASPC..295..477C} and MARX \citep[][]{2012SPIE.8443E..1AD}. 
A PSF model representing the bright nucleus of NGC 7469 is subtracted from the merged data. 
The smoothed soft X-ray zeroth-order image with the PSF model subtracted is shown in Figure~\ref{fig:xray}.

\section{Results}
\label{sec:3}

\subsection{Flux maps of $\rm H\alpha$ and [O~{\sc{iii}}] emission lines}
\label{sec:3.1}

\begin{figure*}
	\gridline{\fig{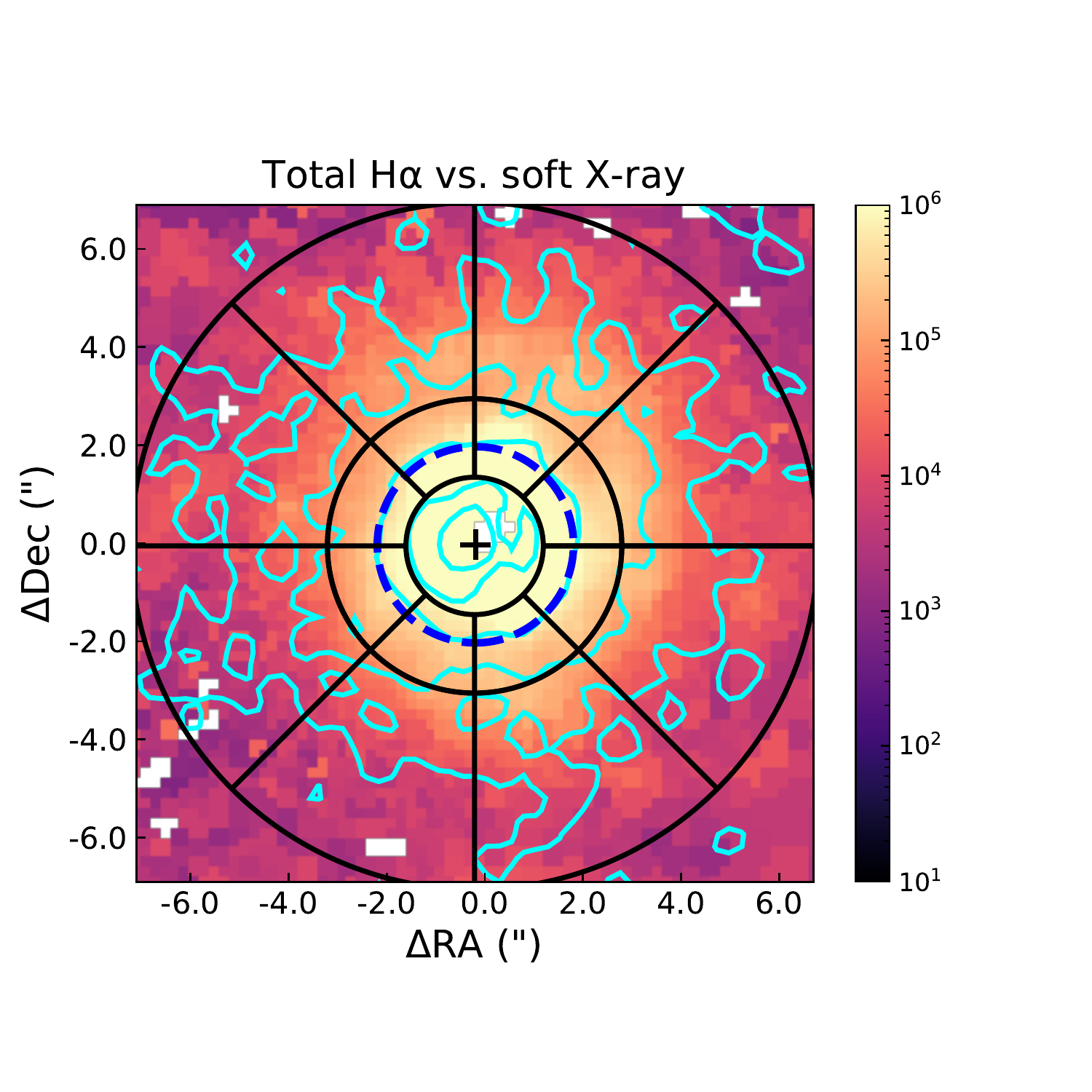}{0.33\textwidth}{(a)}
          \fig{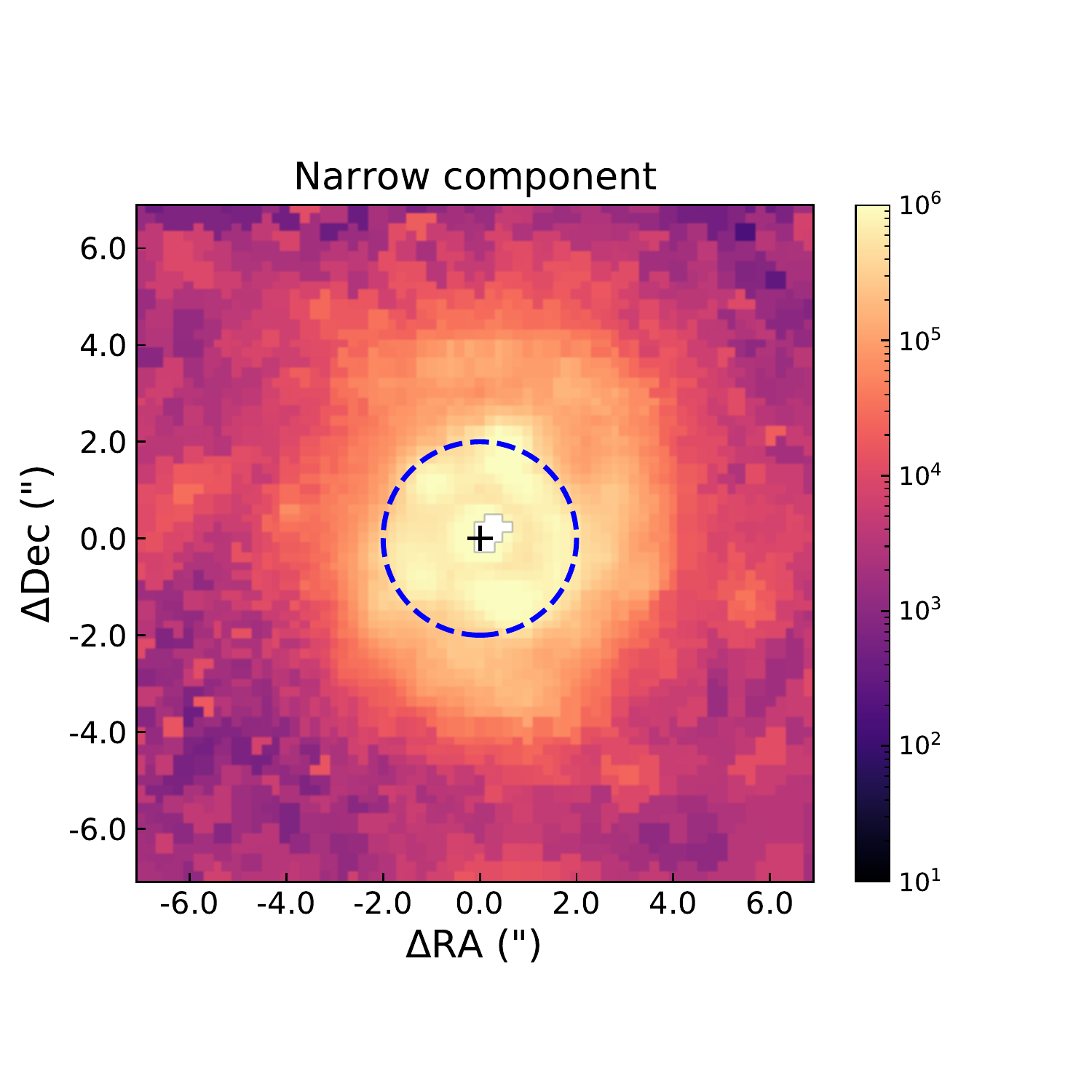}{0.33\textwidth}{(b)}
          \fig{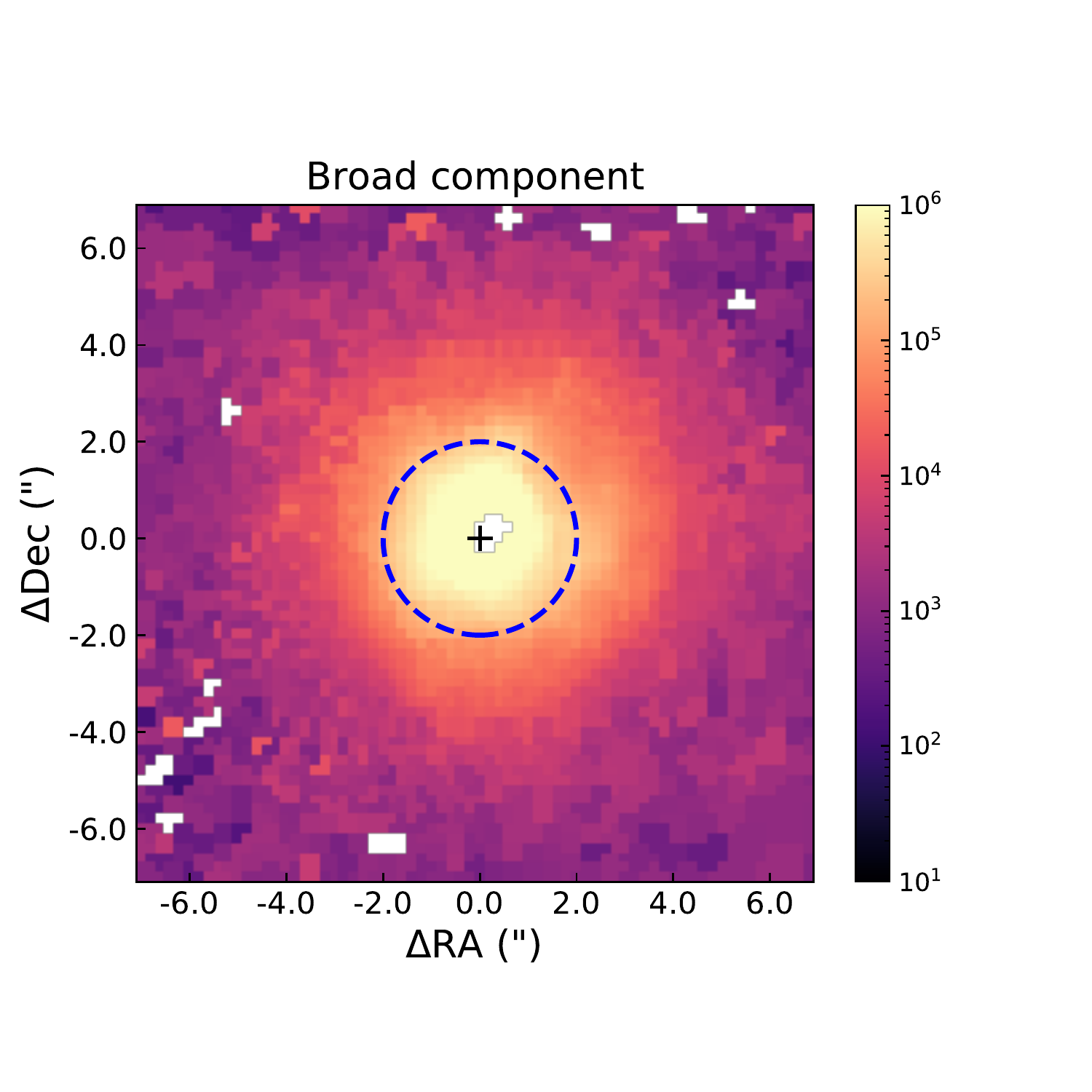}{0.33\textwidth}{(c)}}
	\caption{Flux maps of H$\alpha$. 
	(a): the total flux overlaid with the contours of soft X-ray emission (0.2 - 1 keV, with the nucleus modeled by PSF component subtracted) and the labeled regions in Figure~\ref{fig:xray}. 
	(b): flux map of the narrow component. 
	(c): flux map of the broad component. 
	The color-bar shows the logarithmic flux density with the unit of $\rm 10^{-20}\ erg\ s^{-1}\ cm^{-2}$. 
	The blue dash circle ($r = 2\arcsec$) and black cross indicate the starburst ring and the nucleus, respectively. 
	North is to the top and east is to the left.
	}
	\label{fig:flux_ha}
\end{figure*}

\begin{figure*}
	\gridline{\fig{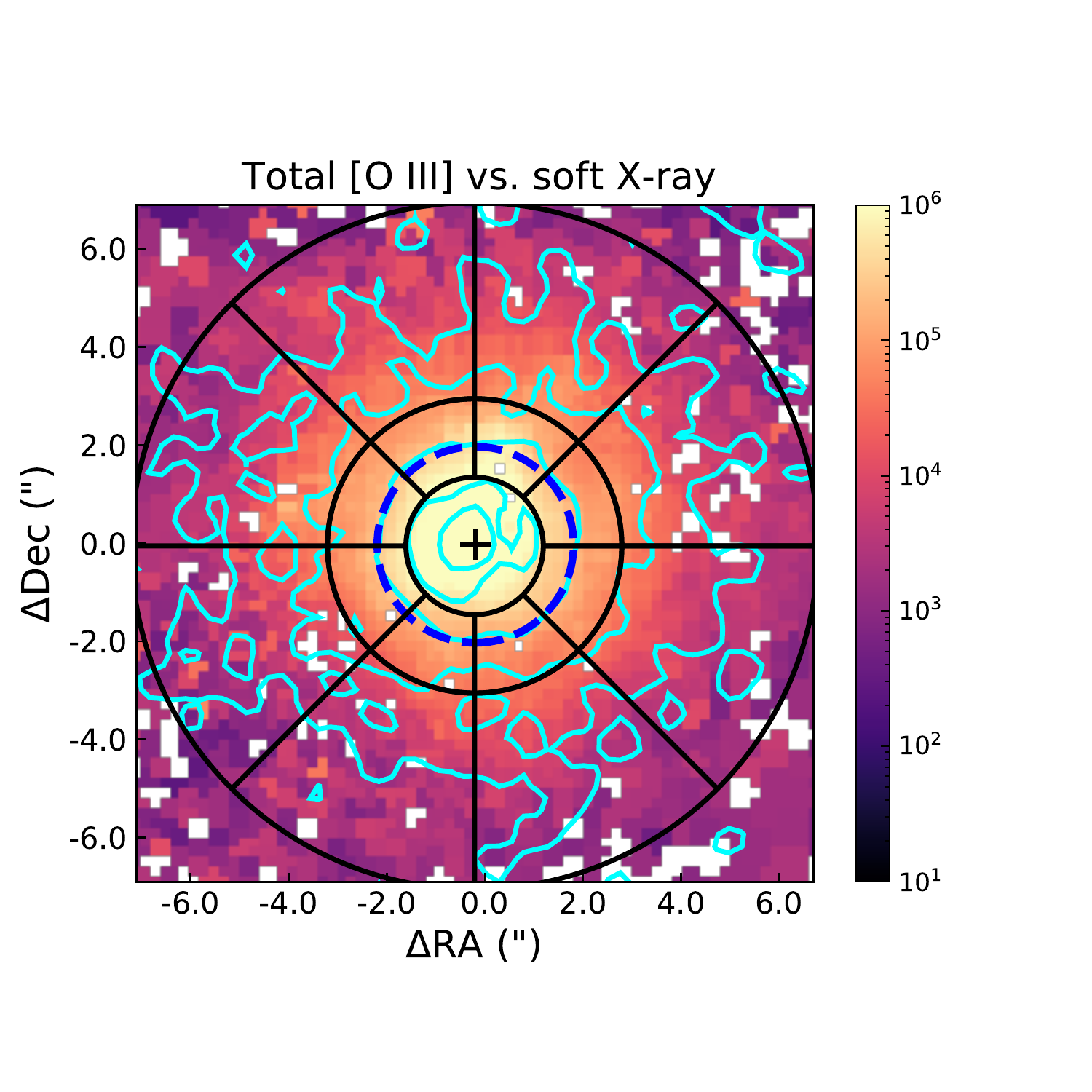}{0.33\textwidth}{(a)}
          \fig{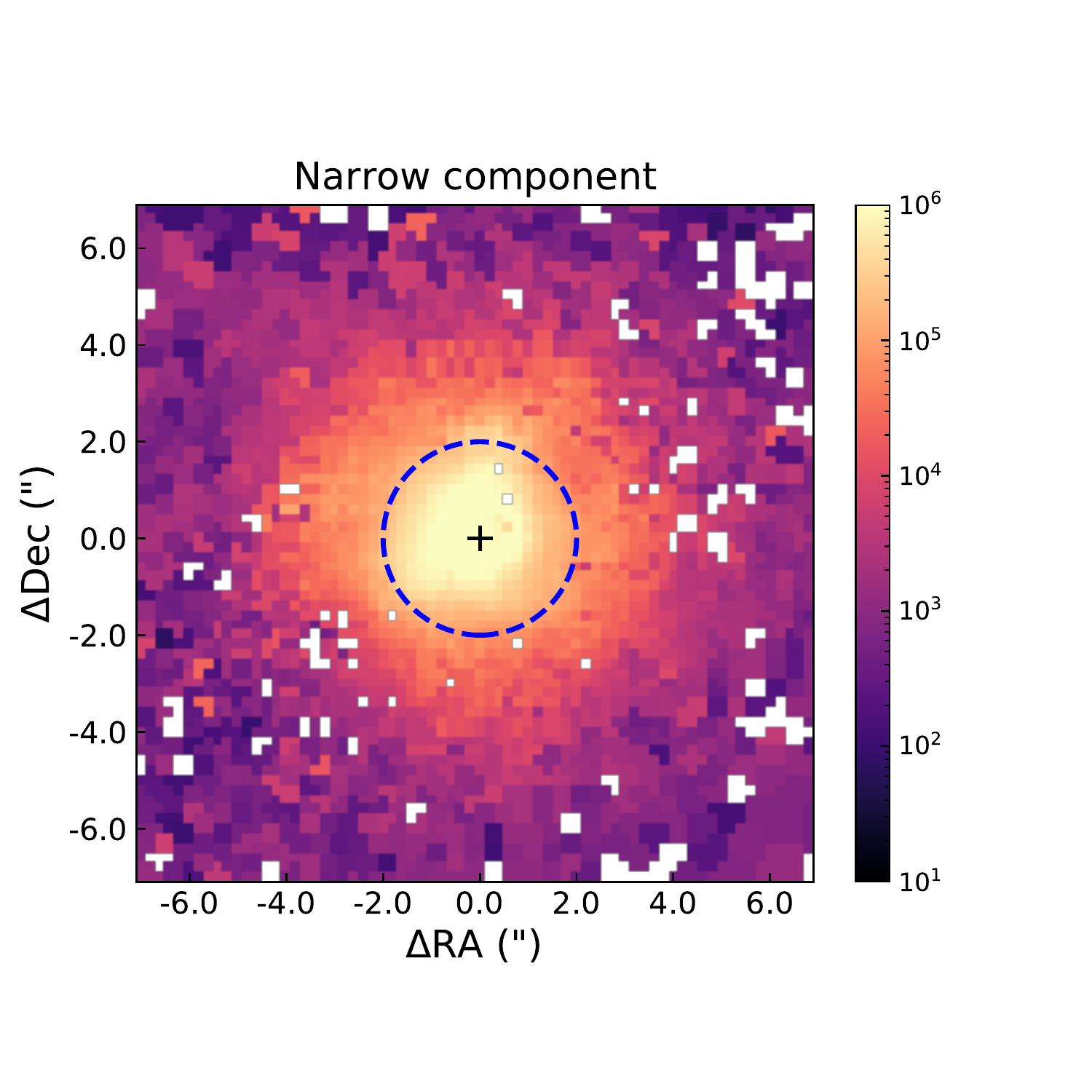}{0.33\textwidth}{(b)}
          \fig{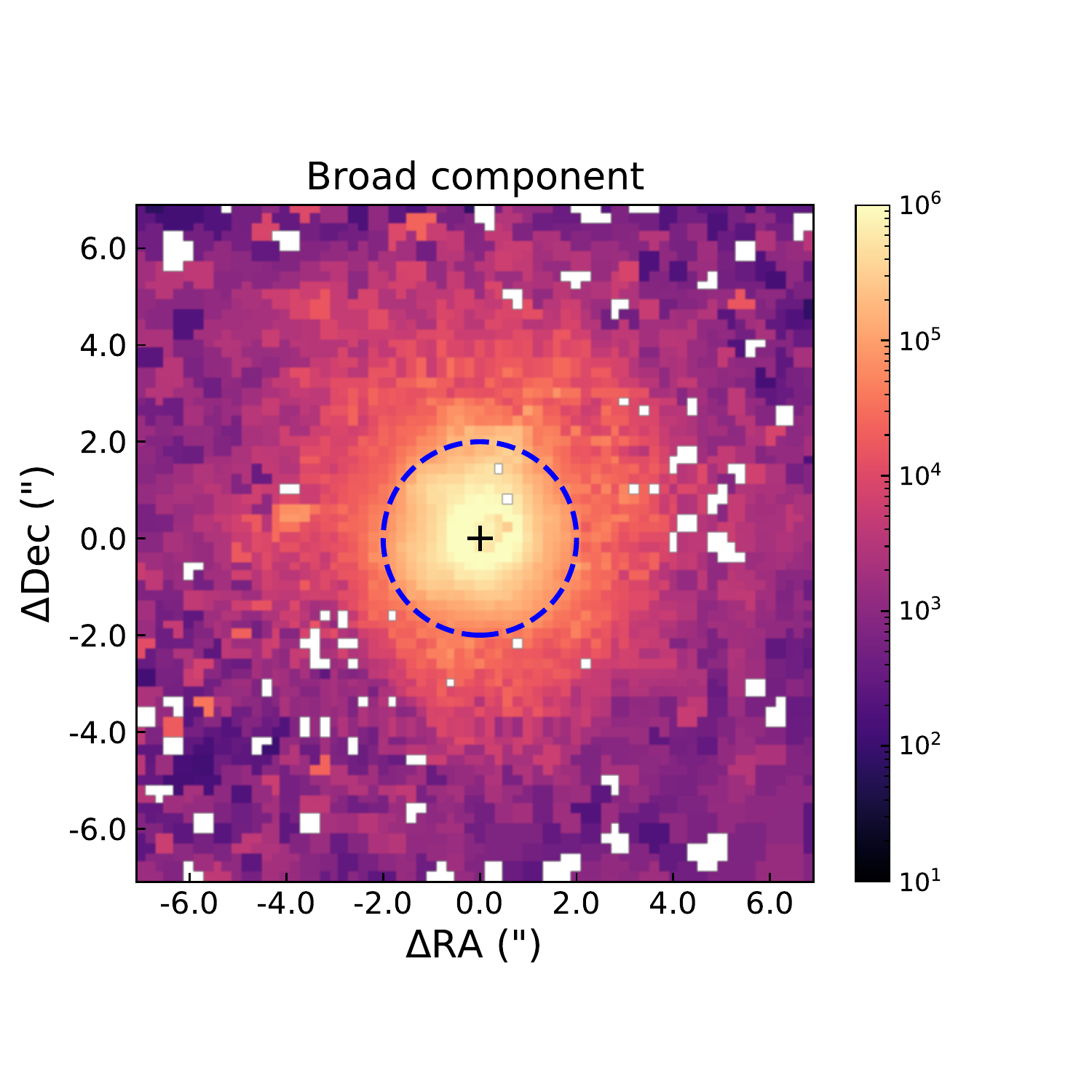}{0.33\textwidth}{(c)}}
	\caption{The same maps as Figure~\ref{fig:flux_ha} but for the [O~{\sc{iii}}]$\lambda$5007 emission line flux. }
	\label{fig:flux_o3}
\end{figure*}

The flux maps of $\rm H\alpha$ from the MUSE data are shown in Figure~\ref{fig:flux_ha}.
In the narrow component flux map of $\rm H\alpha$ (Figure~\ref{fig:flux_ha}b), bright clumpy ionized gas emissions can be found around the nuclear starburst ring, and the spiral arms appear outside the ring. 
The broad component of $\rm H\alpha$ shows a bright nucleus (Figure~\ref{fig:flux_ha}c), which is expected given the line emission is severely contaminated by the broad line region (BLR). 
Several clumps of $\rm H\alpha$ are clearly present along the starburst ring, likely associated with the giant H~{\sc{ii}} complexes. 
The [O~{\sc{iii}}] flux maps (Figure~\ref{fig:flux_o3}) show no clumpy or spiral features.
This may indicate that the [O~{\sc{iii}}] emission is mainly originated from the non-rotational gas instead of the disk.

\begin{figure}
    \includegraphics[width=\columnwidth]{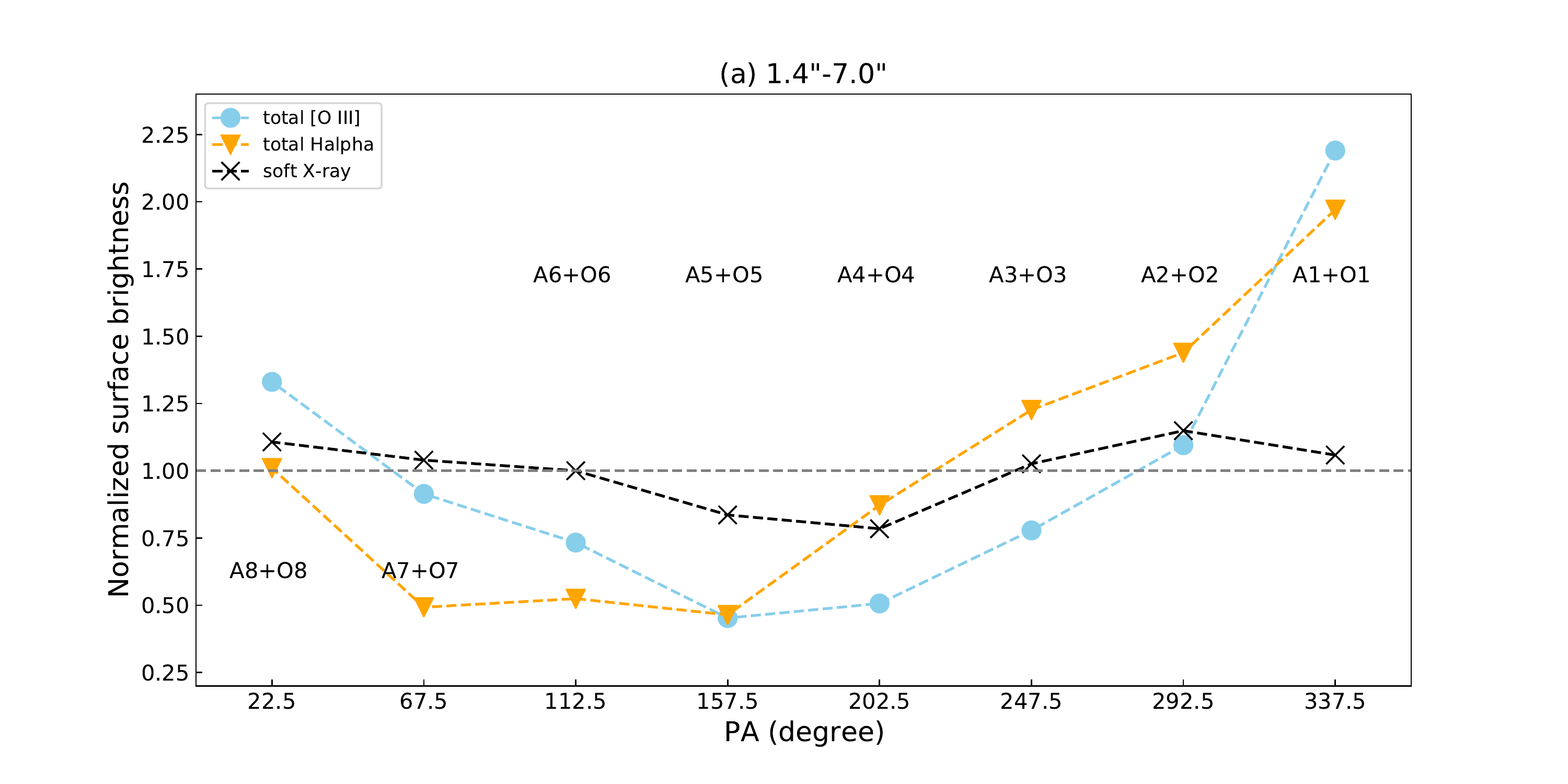}
    \includegraphics[width=\columnwidth]{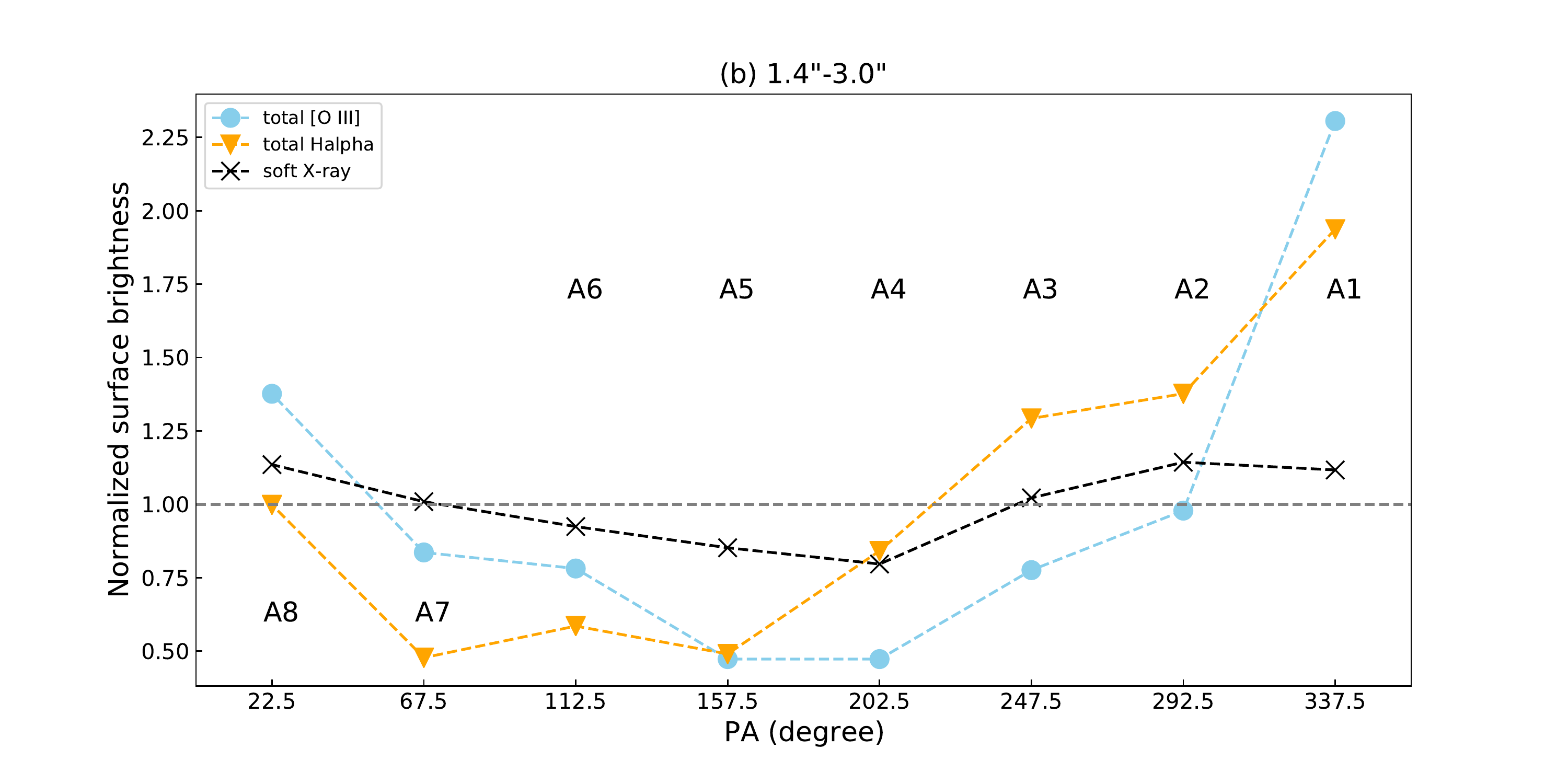}
    \includegraphics[width=\columnwidth]{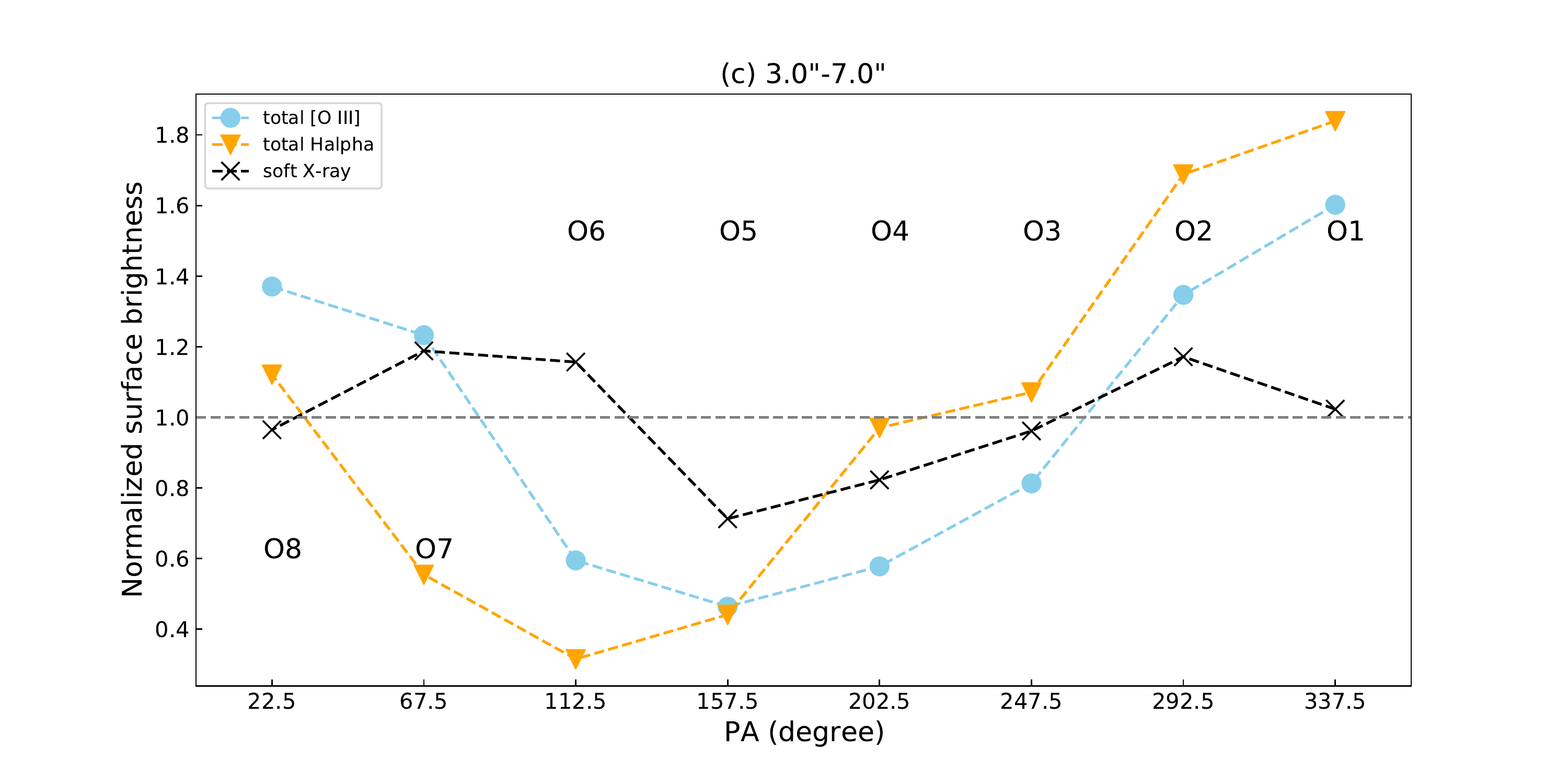}
	\caption{
	Normalized surface brightness (normalized to the mean value) of soft X-ray (black cross and dashed line), $\rm H\alpha$ (orange triangle and dashed line), and [O~{\sc{iii}}]$\lambda$5007 (blue circle and dashed line) in different regions (see Figure~\ref{fig:xray}).
	(a) Regions including the starburst ring and outside sectors (1.\arcsec4--7\arcsec).
	(b) Regions around the starburst ring (1.\arcsec4--3$\arcsec$).
	(c) Regions outside the starburst ring (3$\arcsec$--7$\arcsec$).
	The horizontal gray dashed line denotes the mean surface brightness.
	}
	\label{fig:asymmetry}
\end{figure}

In Figure~\ref{fig:flux_ha}a and~\ref{fig:flux_o3}a, soft X-ray contours are superimposed on the total flux maps of $\rm H\alpha$ and [O~{\sc{iii}}]. 
To quantify the asymmetry of these flux maps, the surface brightness (normalized to the mean value) of different bands in the labeled regions (see Figure~\ref{fig:xray}, \ref{fig:flux_ha}a, and \ref{fig:flux_o3}a) are shown in Figure~\ref{fig:asymmetry}.
Because the nuclear region ($\leq 1.\arcsec4$) is strongly affected by BLR and we mostly focus on the extended emission on a larger scale, this central region is excluded here.
In general, the soft X-ray, $\rm H\alpha$, and [O~{\sc{iii}}] emissions present an asymmetric distribution along the northwest to southeast and the soft X-ray shows less asymmetry than the two optical emission lines in three panels.
As for the $\rm H\alpha$ emission, it tends to be more asymmetric along the east to the west than the soft X-ray and [O~{\sc{iii}}], which can be seen from the enhancement in the west regions (A1, O1, A2, O2, and A3) and the deficiency in the east regions (A5, O5, A6, O6, A7, and O7).
The asymmetry features of the soft X-ray emission around and outside the nuclear region are more consistent with the [O~{\sc{iii}}] emission than $\rm H\alpha$.

\subsection{Kinematics maps of ionized gas}
\label{sec:3.2}

\begin{figure*}
	\gridline{\fig{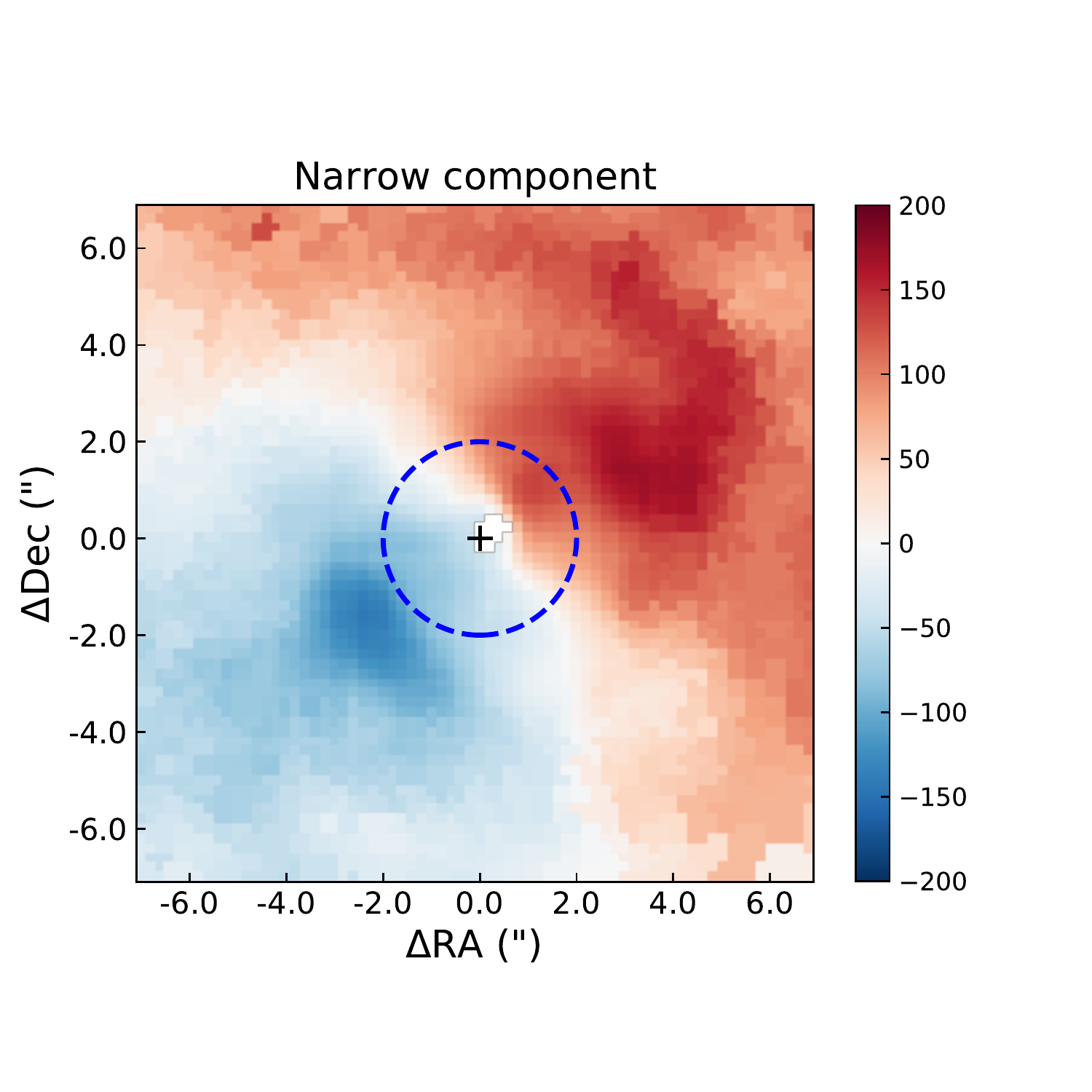}{0.45\textwidth}{(a)}
          \fig{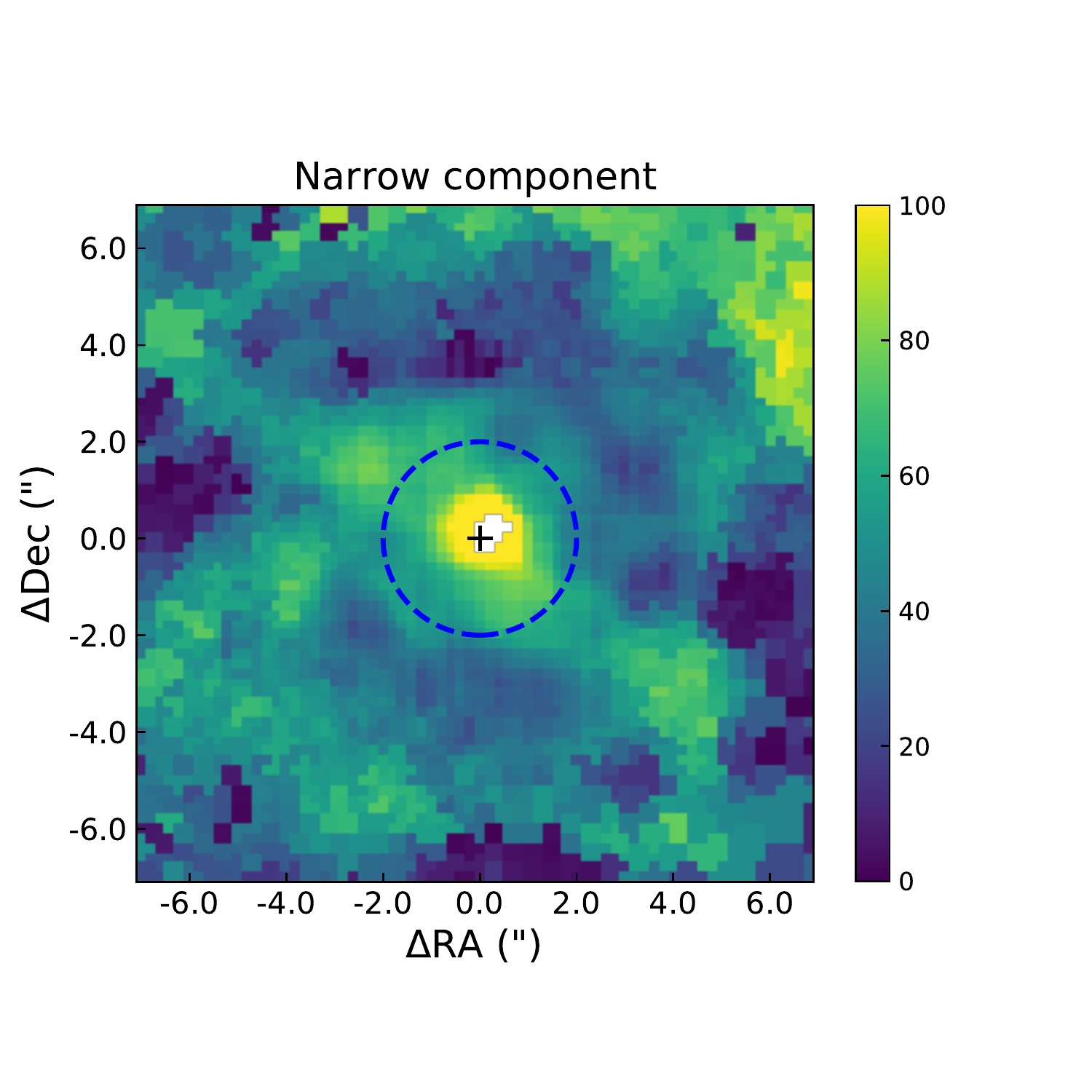}{0.45\textwidth}{(b)}}
          
    \gridline{\fig{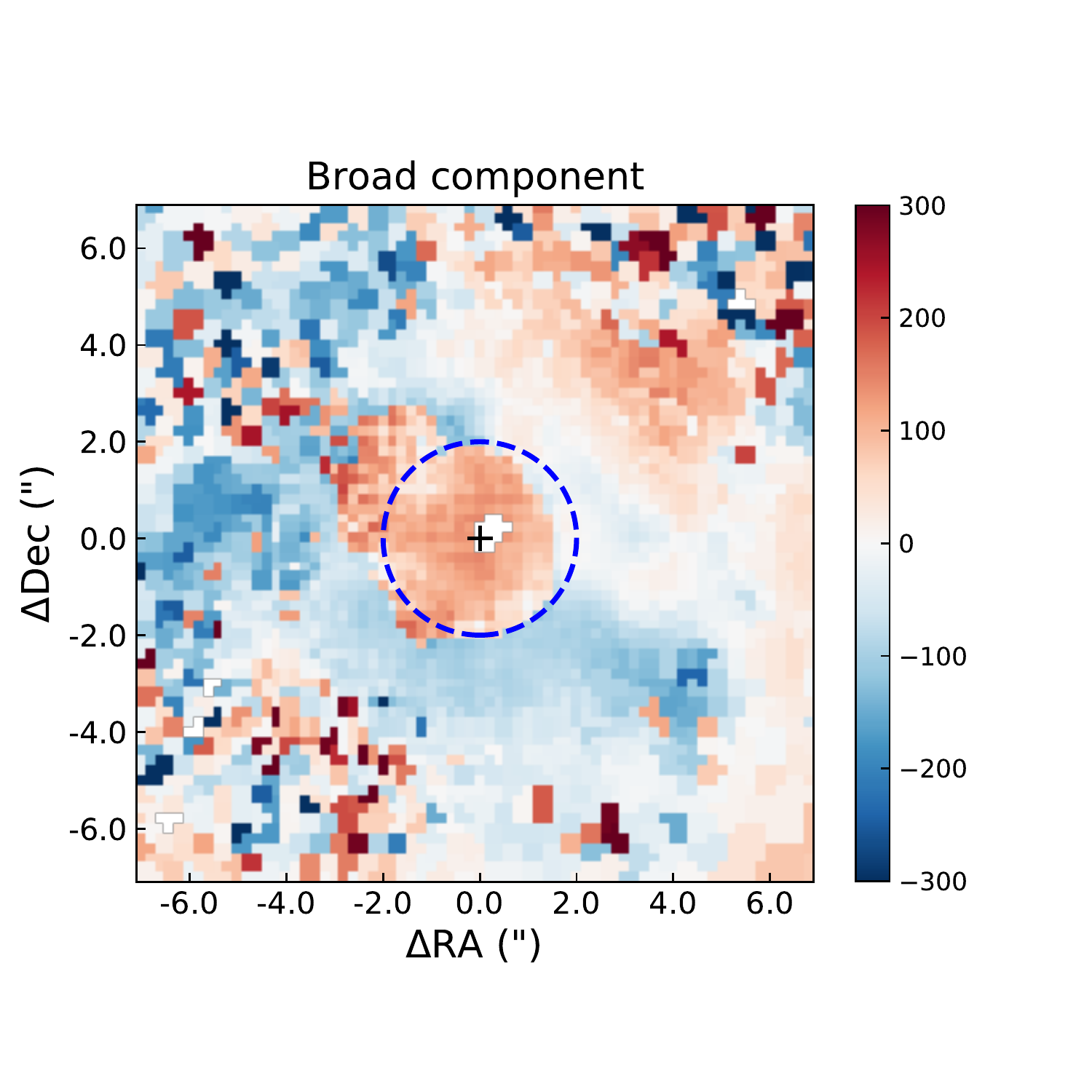}{0.45\textwidth}{(c)}
          \fig{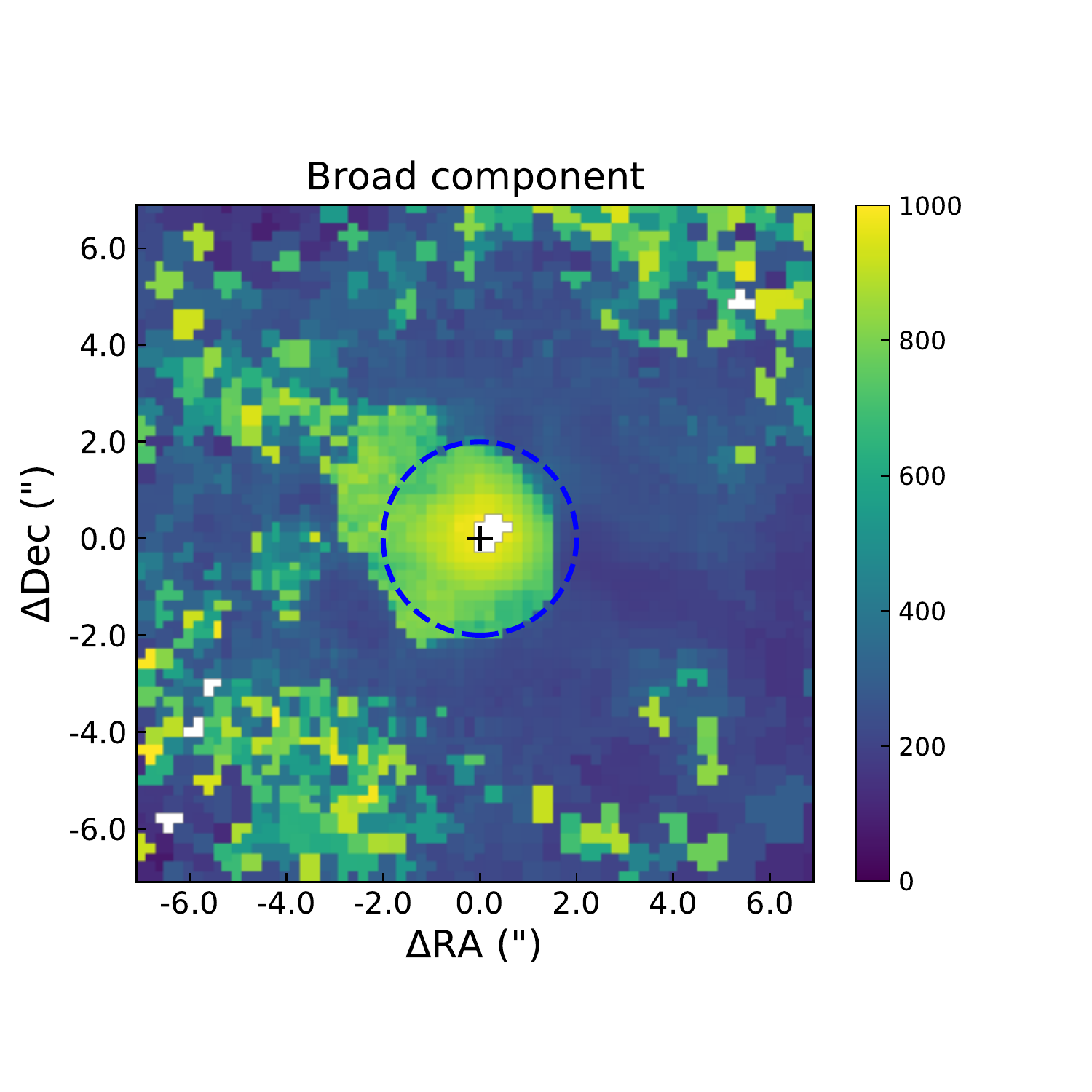}{0.45\textwidth}{(d)}}
    \caption{Velocity and velocity dispersion maps of H$\alpha$ emission line with the unit of $\rm km s^{-1}$. 
    (a) velocity map of the narrow component. 
    (b) velocity dispersion map of the narrow component.
    (c) velocity map of the broad component.  
    (d) velocity dispersion map of the broad component. 
    North is to the top and east is to the left.
    }
    \label{fig:kin_ha}
\end{figure*}

The kinematics maps of $\rm H\alpha$ ionized gas are shown in Figure~\ref{fig:kin_ha}. 
Note that $\rm H\alpha$, $\rm H\beta$, and [N~{\sc{ii}}]$\lambda\lambda$6548,6583 were fitted with the same velocity and velocity dispersion.
For simplicity, the kinematics maps of $\rm H\alpha$ in fact represent the kinematics for these emission lines. 
Because the kinematics of [S~{\sc{ii}}]$\lambda\lambda$6716,6731 (fitted with a single component) is very similar to the narrow component of $\rm H\alpha$, we do not present the velocity and velocity dispersion maps of [S~{\sc{ii}}]$\lambda\lambda$6716,6731 here.

\begin{figure*}
	\gridline{\fig{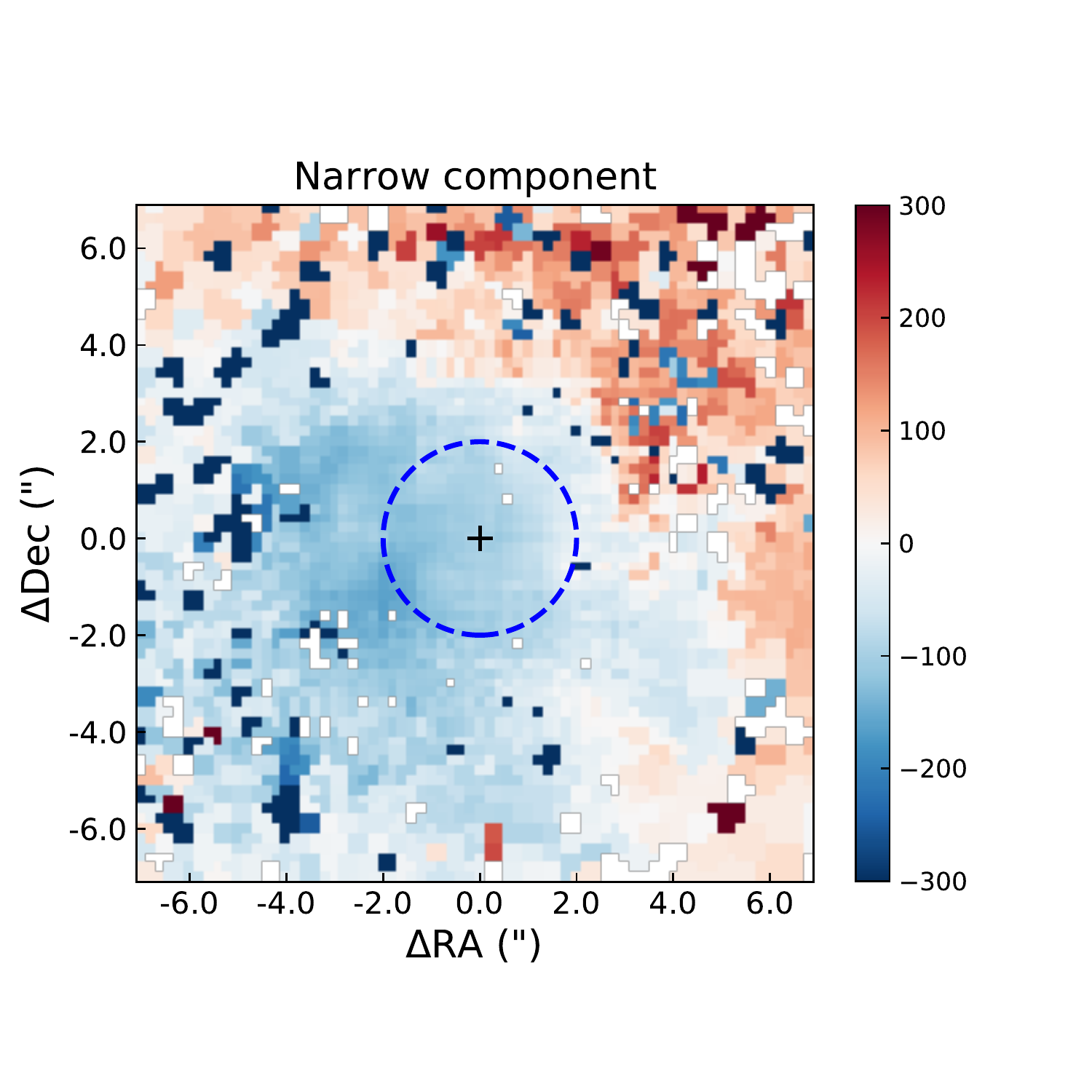}{0.45\textwidth}{(a)}
          \fig{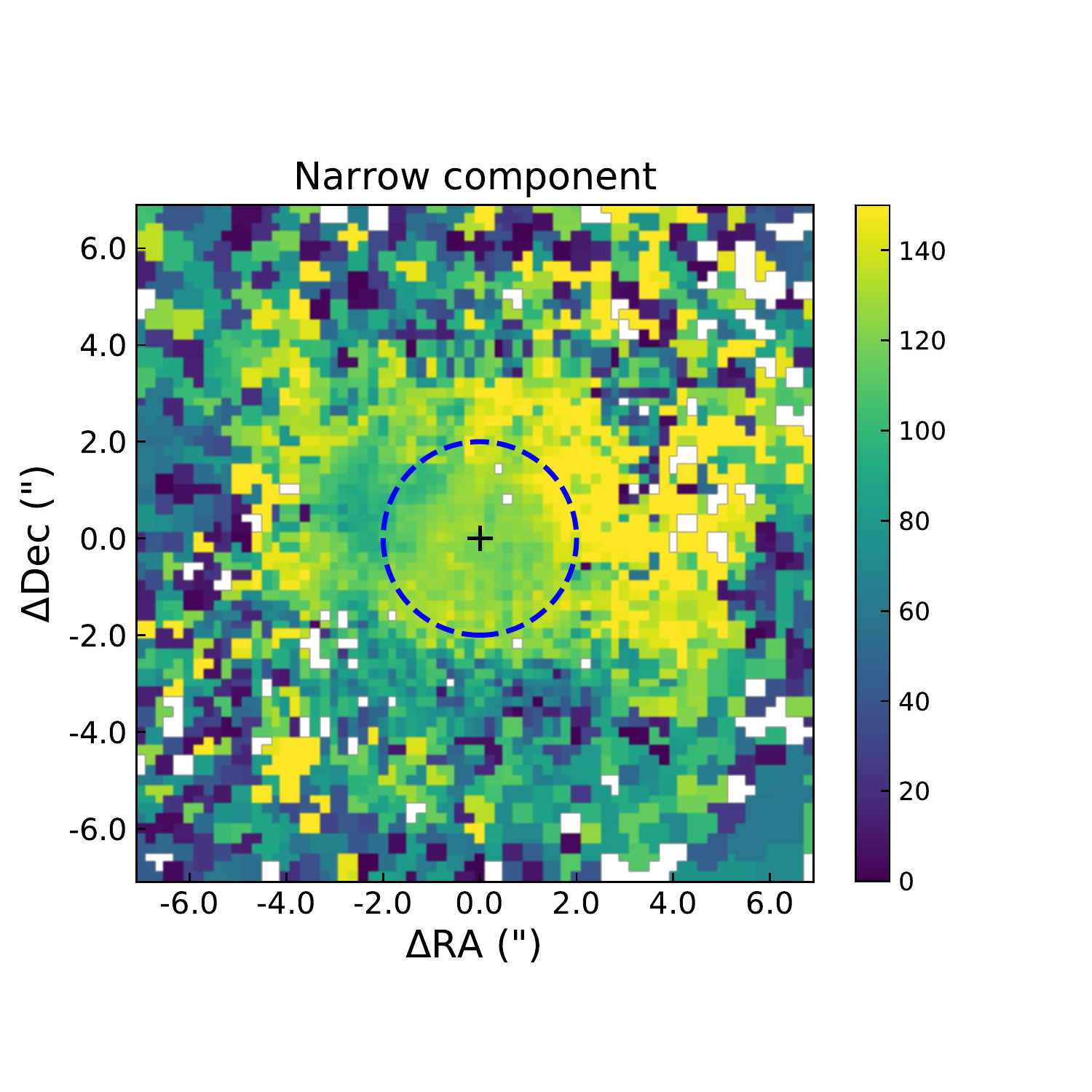}{0.45\textwidth}{(b)}}
          
    \gridline{\fig{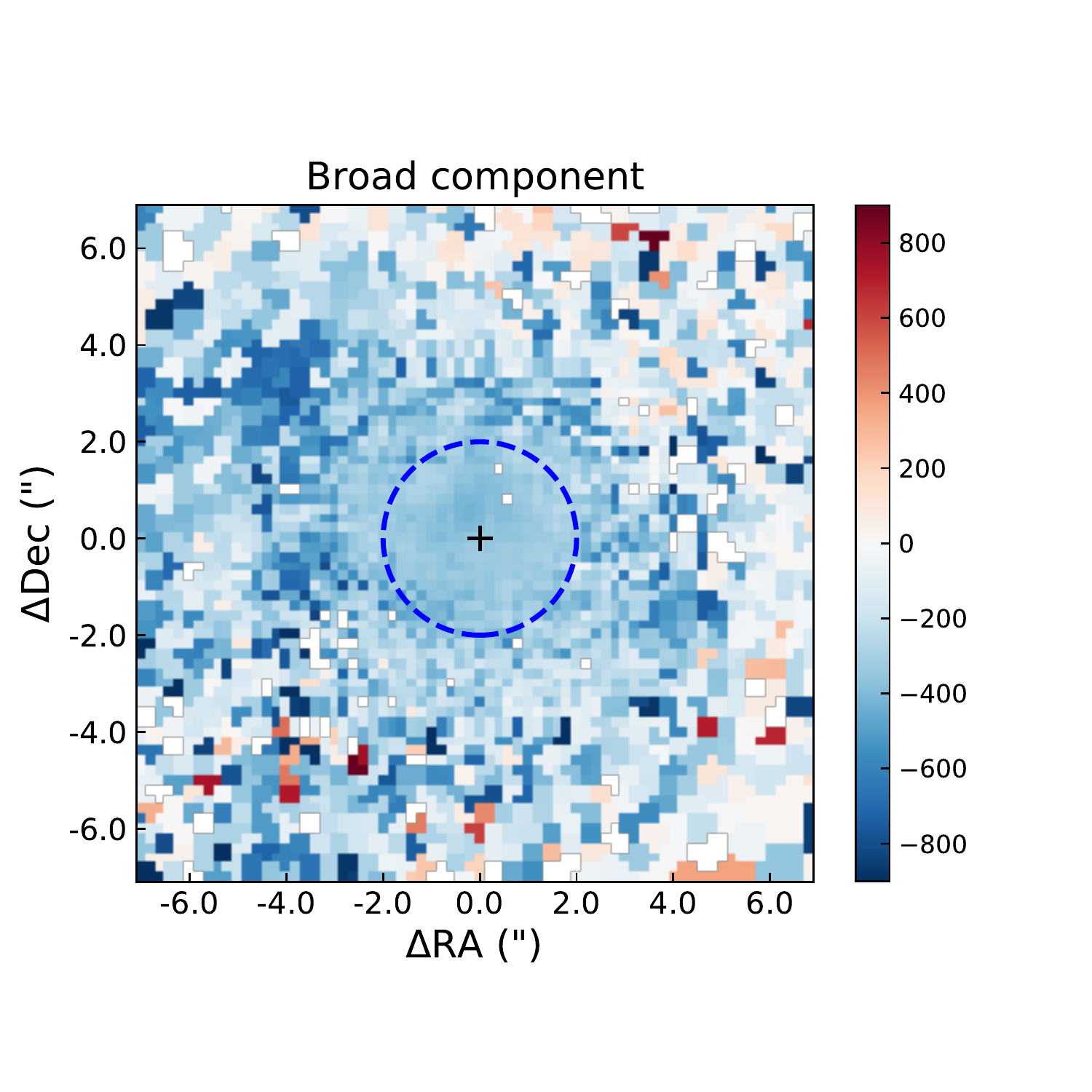}{0.45\textwidth}{(c)}
          \fig{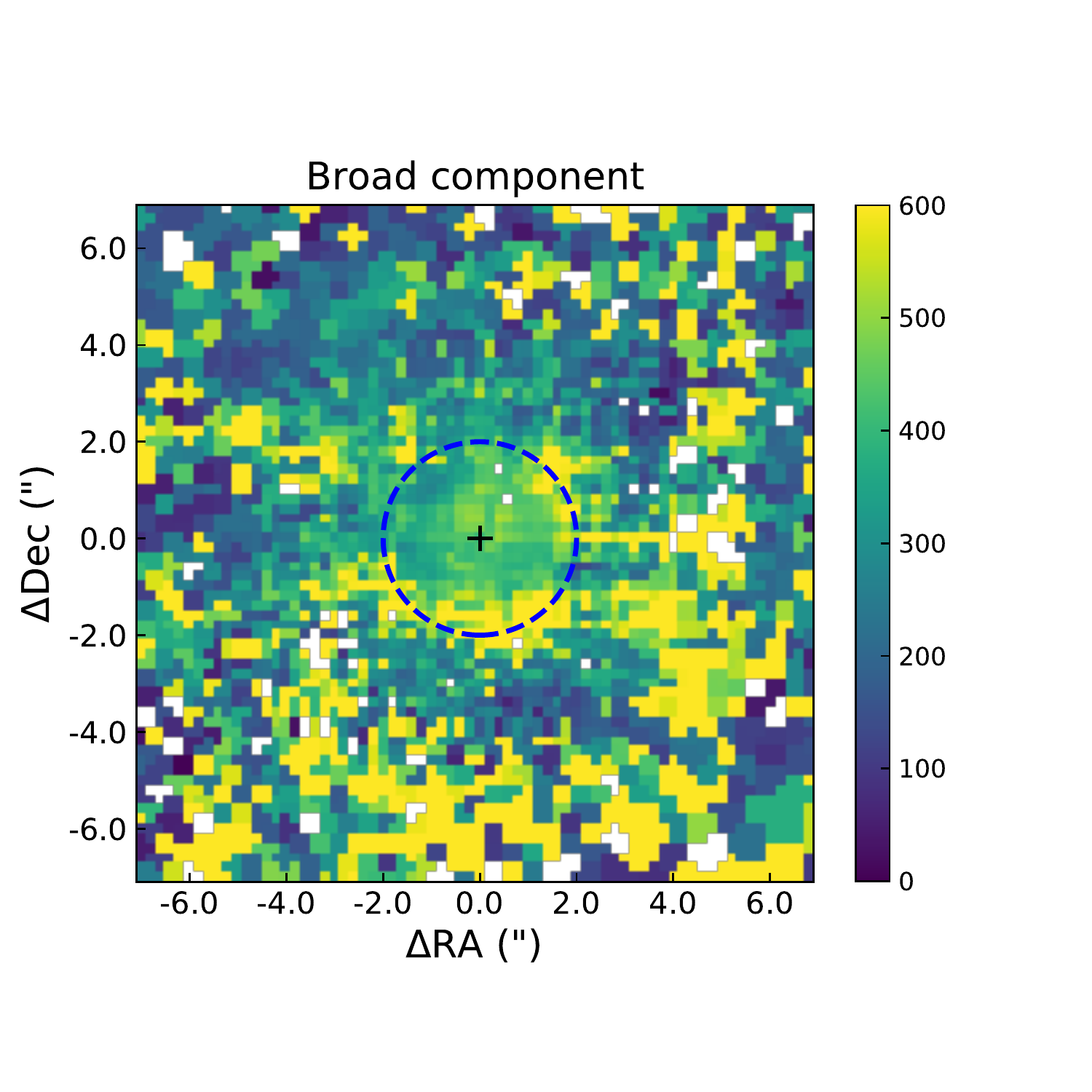}{0.45\textwidth}{(d)}}
    \caption{Same maps as Figure~\ref{fig:kin_ha} but for [O~{\sc{iii}}]$\lambda$5007.}
    \label{fig:kin_o3}
\end{figure*}

The narrow component of $\rm H\alpha$ is dominated by the ionized gas disk due to the rotational pattern shown by the velocity and velocity dispersion maps (Figures~\ref{fig:kin_ha}a and~\ref{fig:kin_ha}b). We verified that the stellar kinematics is very similar to that of the narrow component of H$\alpha$ by fitting the stellar continuum in individual spaxels. 
Previous studies have also established that the kinematics maps of narrow components of $\rm H\alpha$ are dominated by the gravitational potential of galaxy \citep[e.g.][]{2016ApJ...819..148K}. 
For the broad component, complex patterns are present in the kinematics maps of $\rm H\alpha$ (Figure~\ref{fig:kin_ha}c). 
For [O~{\sc{iii}}] the velocity map (Figure~\ref{fig:kin_o3}) of the narrow component cannot be solely explained by a rotating gas disk. 
However, the velocity map of [O~{\sc{iii}}] shows mostly blueshifted features across the FoV. 
Comparing narrow and broad components of $\rm H\alpha$ and [O~{\sc{iii}}], the corresponding relationships between them cannot be directly obtained. 
A possible reason is that [O~{\sc{iii}}] doublet and Balmer emission lines are dominated by different ionized gas component. 
Balmer emission lines is sensitive to rotational disk and low-velocity ionized outflows, while [O~{\sc{iii}}] traces NLR and relatively high-velocity ionized outflows. 
To compare and quantify the kinematics of Balmer emission lines and [O~{\sc{iii}}] simultaneously, the non-parametric kinematics maps are further analyzed.

\begin{figure*}
	\gridline{\fig{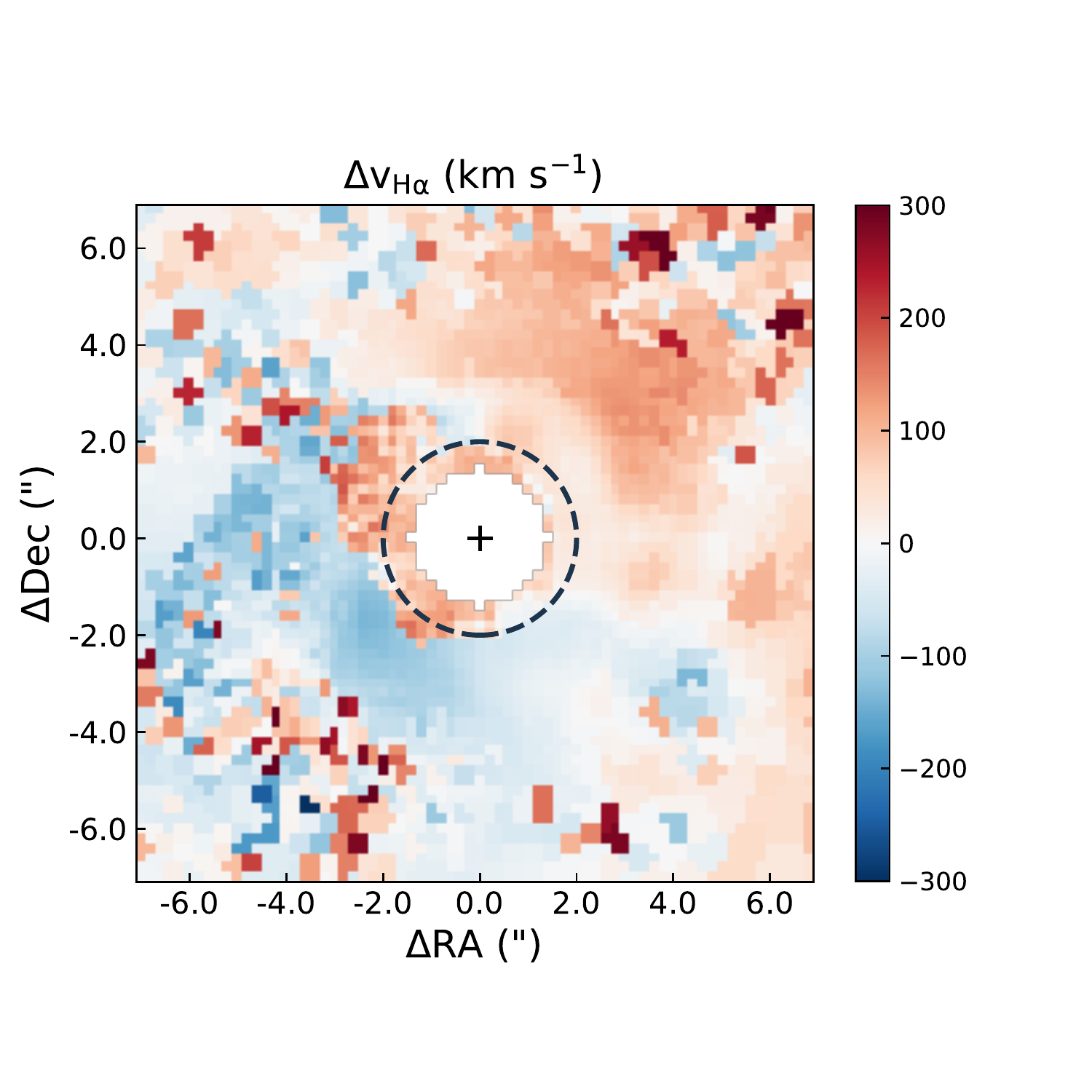}{0.33\textwidth}{(a)}
          \fig{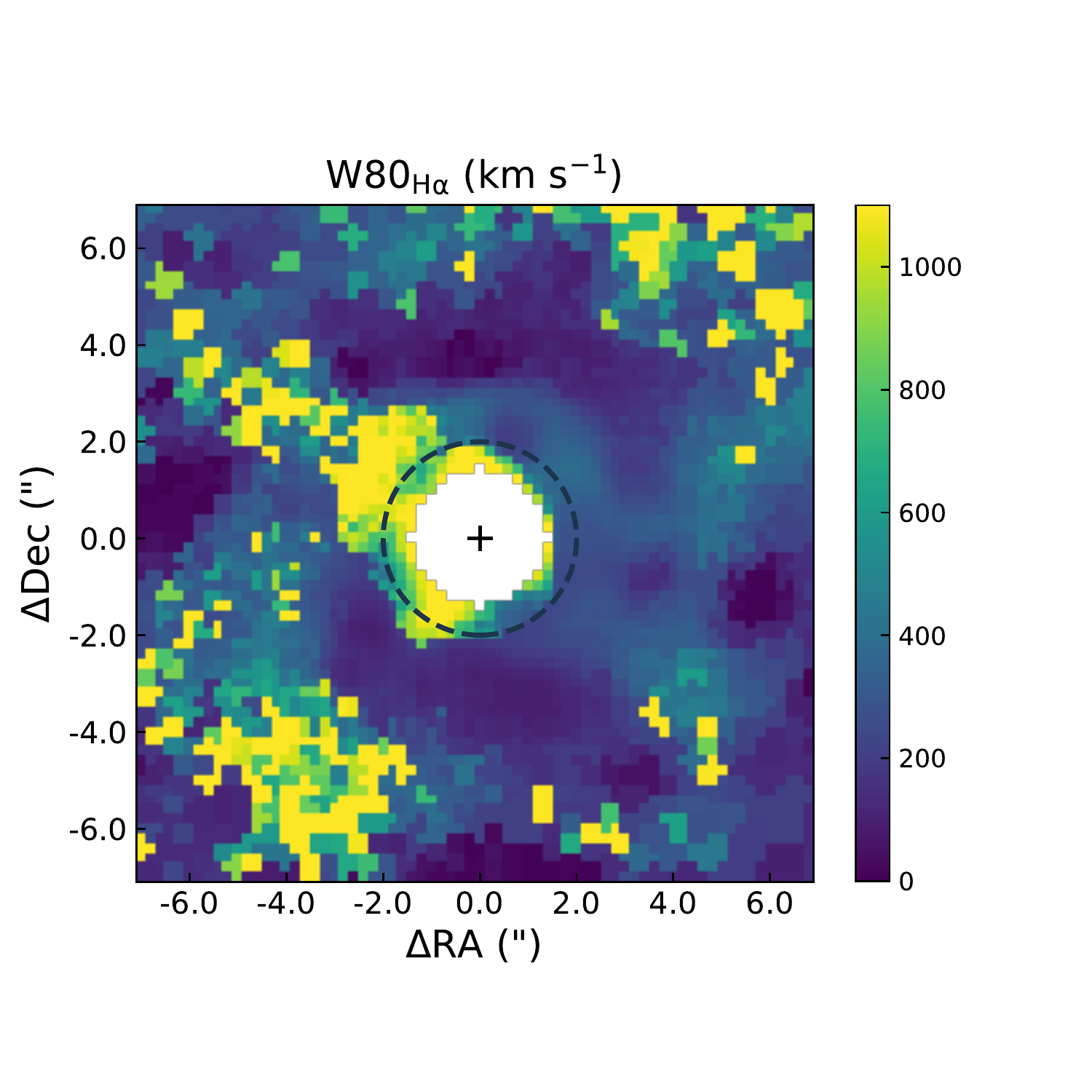}{0.33\textwidth}{(b)}
          \fig{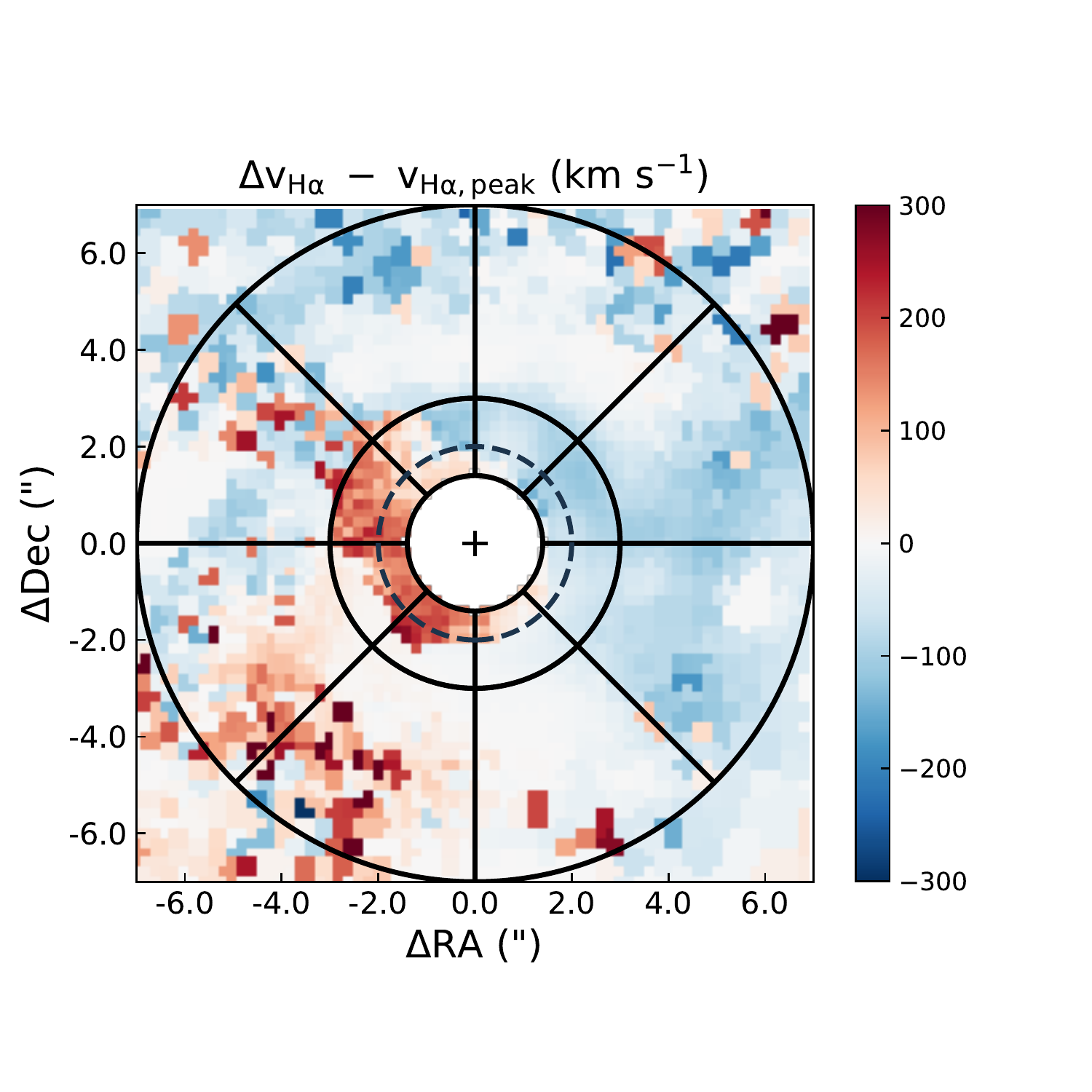}{0.33\textwidth}{(c)}}
	\caption{Non-parametric kinematics maps of H$\rm \alpha$ emission line. 
	(a) the $\Delta v_{\rm H\alpha}$ map. 
	(b) the $\rm W80_{H\alpha}$ map. 
	(c) the map of $\Delta v_{\rm H\alpha}-v_{\rm H\alpha,peak}$. 
	Except for the $v_{\rm H\alpha,peak}$, other maps all exclude the nuclear region ($<1.\arcsec4$) for the influence of BLR.
	North is to the top and east is to the left.
	}
	\label{fig:np_kin_ha}
\end{figure*}

\begin{figure*}[htbp!]
	\gridline{\fig{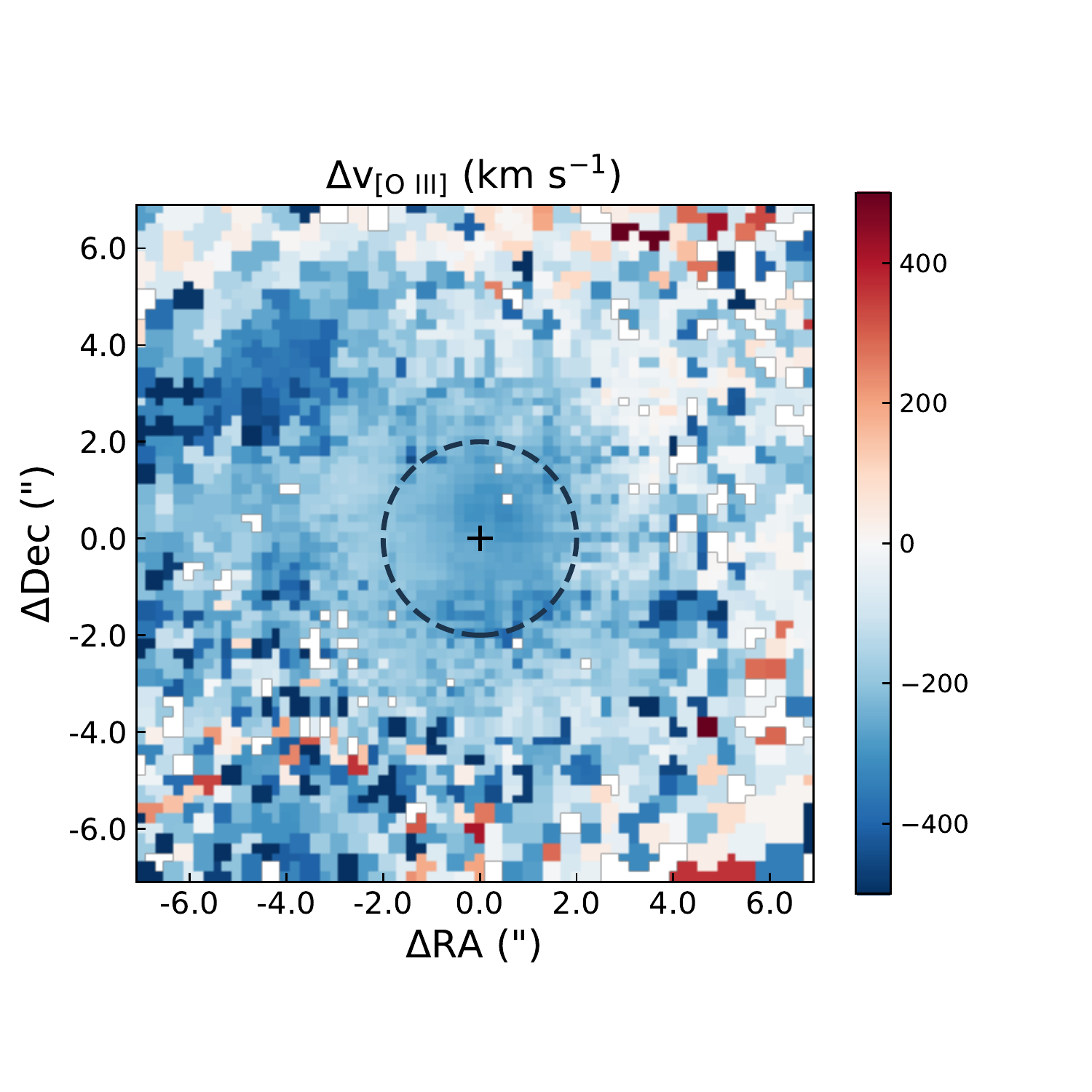}{0.33\textwidth}{(a)}
          \fig{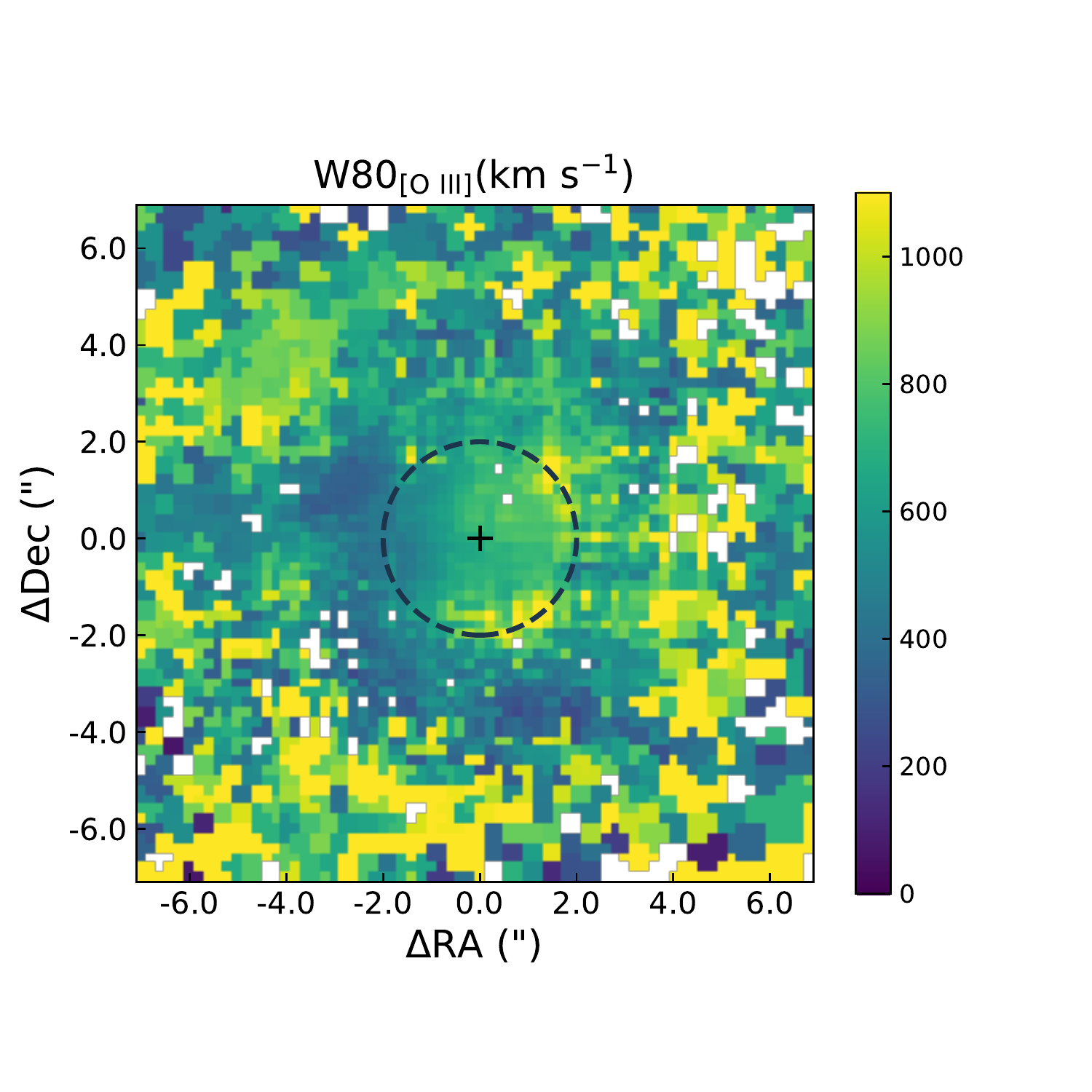}{0.33\textwidth}{(b)}
          \fig{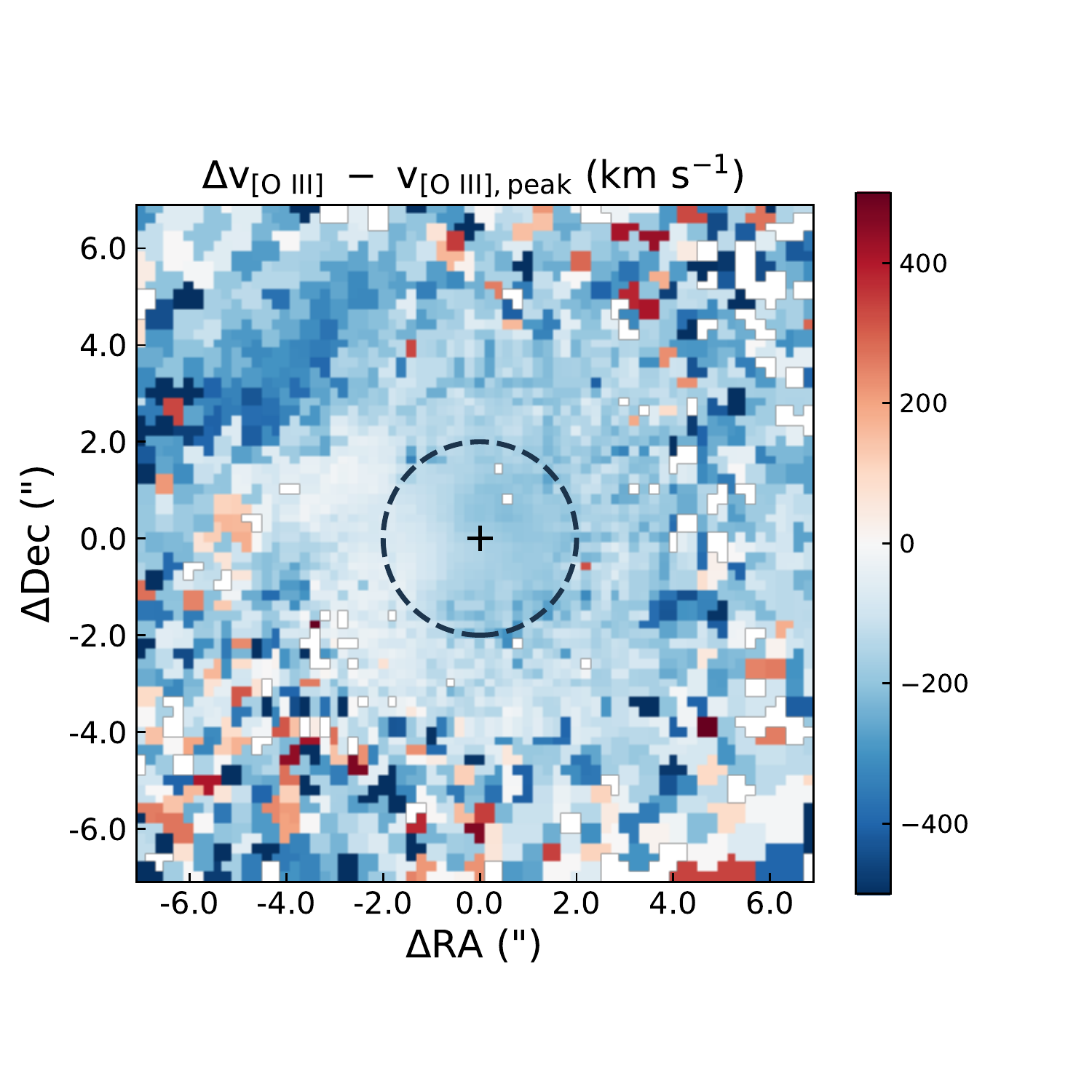}{0.33\textwidth}{(c)}}
          
    \gridline{
          \fig{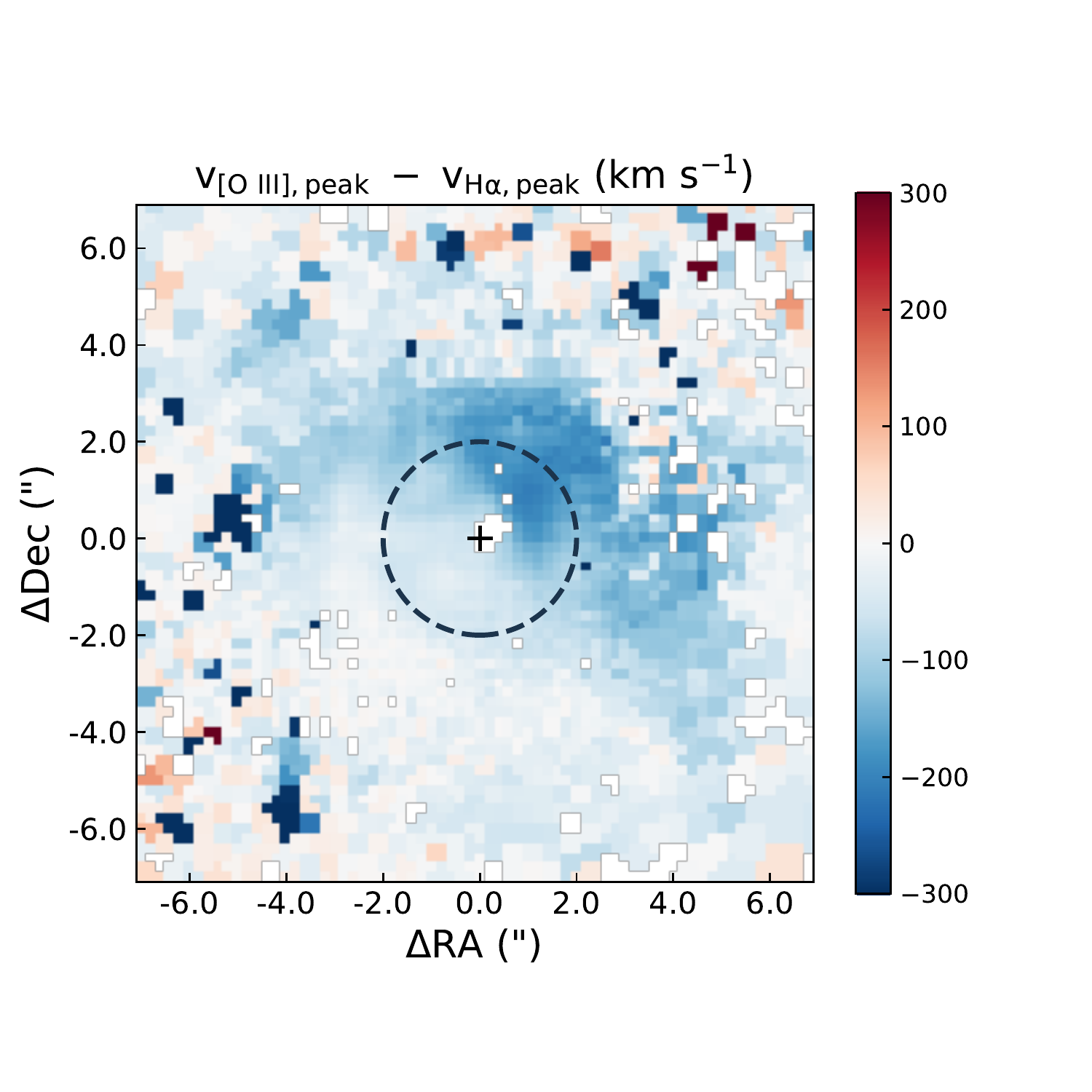}{0.45\textwidth}{(d)}
          \fig{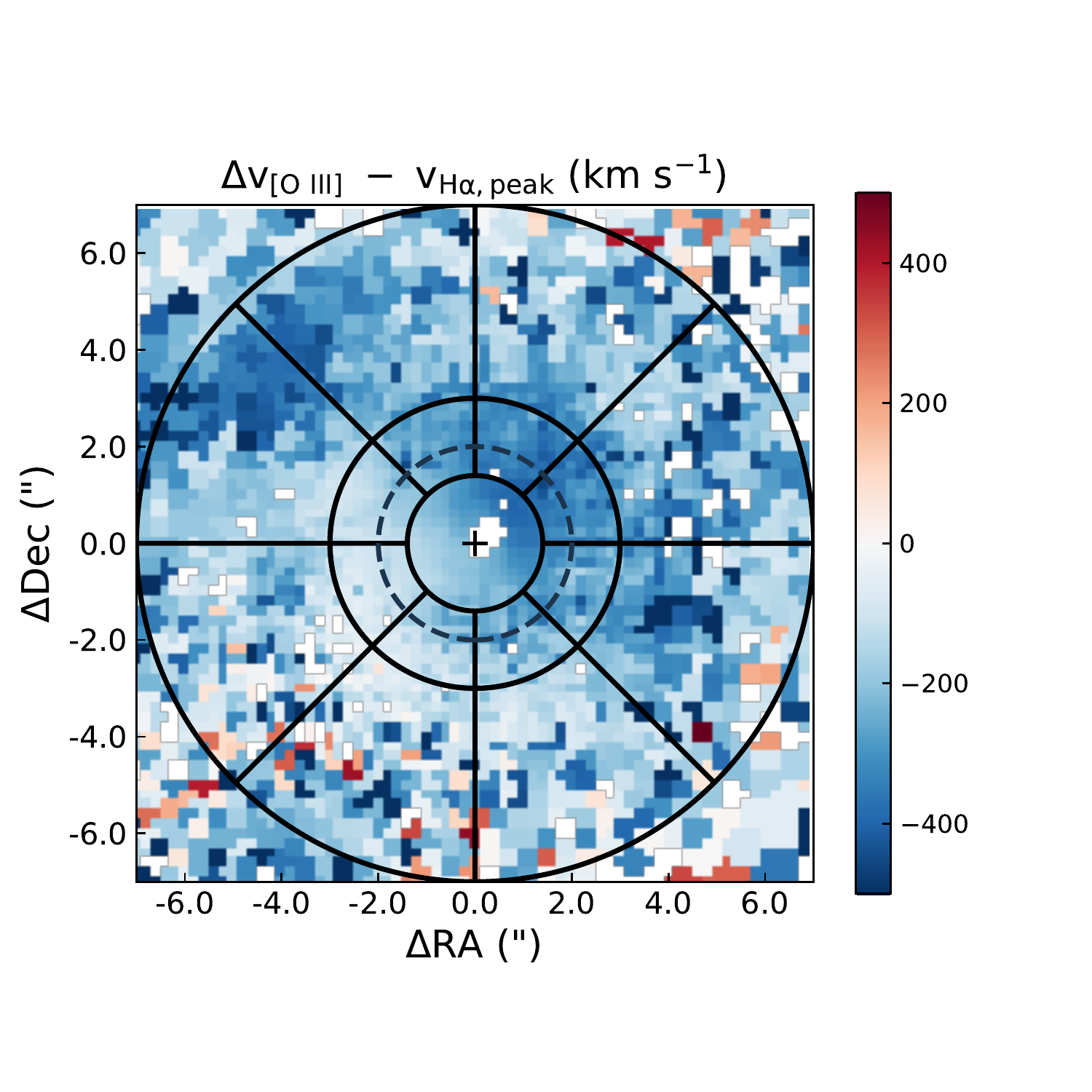}{0.45\textwidth}{(e)}
          }
	\caption{Non-parametric kinematics maps of [O~{\sc{iii}}]$\lambda$5007 emission line. 
	(a) the $\Delta v_{\rm [O\, III]}$ map. 
	(b) the $\rm W80_{[O\, III]}$ map. 
	(c) the map of $\Delta v_{\rm [O\, III]}-v_{\rm [O\, III],peak}$.
	(d) the map of $v_{\rm [O\, III],peak}-v_{\rm H\alpha,peak}$. 
	(e) the map of $\Delta v_{\rm [O\, III]}-v_{\rm H\alpha,peak}$. 
	North is to the top and east is to the left.
	}
	\label{fig:np_kin_o3}
\end{figure*}

Figures~\ref{fig:np_kin_ha} and \ref{fig:np_kin_o3} show the non-parametric kinematics maps of $\rm H\alpha$ and [O~{\sc{iii}}], respectively. 
Note that the nuclear region ($<1.\arcsec4$) are excluded in three panels ($\Delta v_{\rm H\alpha}$, $\rm W80_{H\alpha}$, and $\Delta v_{\rm H\alpha} - v_{\rm H\alpha,peak}$) due to the contamination from the bright BLR. 
The $v_{\rm peak}$ maps of $\rm H\alpha$ and [O~{\sc{iii}}] appear similar to their velocity maps of narrow component, hence we do not show them here.
Theoretically, the velocity offset $\Delta v$ is affected by two physical components: a rotational gas disk and non-rotational gas component. 
If an emission line is dominated by a rotational component and can be fitted with a single Gaussian model well, the velocity offset $\Delta v$ will be equal to the $v_{\rm peak}$ and reflect the rotational velocity.
When there exist another component (i.e. non-rotational component), an extra velocity offset from this component will be included in the $\Delta v$.
For example, the $\Delta v_{\rm H\alpha}$ map (Figures~\ref{fig:np_kin_ha}a) shows a rotational feature along the northwest to southeast, whereas this feature disappears in the $\Delta v_{\rm H\alpha} - v_{\rm H\alpha,peak}$ map (Figures~\ref{fig:np_kin_ha}c).
Therefore, the $\Delta v_{\rm H\alpha} - v_{\rm H\alpha,peak}$ map can represent the velocity offsets originated from the non-rotational gas component.

The $\Delta v_{\rm H\alpha} - v_{\rm H\alpha,peak}$ map shows complex and intriguing patterns. 
As the $v_{\rm H\alpha,peak}$ map is dominated by the rotational disk, this residual implies significant presence of the non-rotational component.
In the $\Delta v_{\rm H\alpha} - v_{\rm H\alpha,peak}$ map, a remarkable redshifted structure extending from the A7 region to the O7.
In the corresponding regions of $\rm W80_{H\alpha}$ map, high values of W80 is suggestive that this redshifted feature is an inflow-like structure.  
Another distinct feature in this map is located at the southeast corner of the FoV (O5 and O6 regions) with redshifted velocity and high W80 value, which may also resemble an inflowing gas stream. 
In addition, the southeast redshifted structure appears in contact with a blueshifted feature in the O6 region.  
The presence of $\rm H\alpha$ outflow is evident as shown in the A1, A2, O2, A3, and O3 regions, with significantly blueshifted emission ($\Delta v_{\rm H\alpha} - v_{\rm H\alpha,peak}$).
In the corresponding regions of $\rm W80_{H\alpha}$ map, relatively high values of W80 are also consistent with an outflow feature.

As for the [O~{\sc{iii}}] maps (Figure~\ref{fig:np_kin_o3}), the $\Delta v_{\rm [O\, III]}$ map shows mostly blueshifted velocity offsets across the FoV.
The $\Delta v_{\rm [O\, III]} - v_{\rm [O\, III],peak}$ and the $\Delta v_{\rm [O\, III]} - v_{\rm H\alpha,peak}$ map are shown in Figures~\ref{fig:np_kin_o3}c and~\ref{fig:np_kin_o3}e. 
The former presents less blueshifted than the later especially in the A1, A2, A3, and A8 regions.
The $v_{\rm [O\, III],peak}$ map cannot be explained only by the rotational ionized gas disk, with an excess of significant blueshifted velocity especially in the A1, A2, A3, and A8 regions (Figure~\ref{fig:np_kin_o3}d), which indicates in these regions the non-rotational component is more significant.
Because the $v_{\rm H\alpha,peak}$ map closely resembles a rotational disk than $v_{\rm [O\, III],peak}$, we use $\Delta v_{\rm [O\, III]} - v_{\rm H\alpha,peak}$ to represent the ionized gas outflow component traced by [O~{\sc{iii}}]. 
After subtracting the rotational component ($v_{\rm H\alpha,peak}$) from the $\Delta v_{\rm [O\, III]}$ map, the remaining shows blueshifted velocity offsets across the FoV (Figure~\ref{fig:np_kin_o3}e), especially in the western regions around the starburst ring (A1, A2, A3) and the regions outside the ring (O1-O8). 
The $\rm W80_{O\, III}$ map (Figures~\ref{fig:np_kin_o3}b) also shows good correspondence with the $\Delta v_{\rm [O\, III]} - v_{\rm H\alpha,peak}$ map, indicating the $\Delta v_{\rm [O\, III]} - v_{\rm H\alpha,peak}$ map is dominated by ionized gas outflows.

\subsection{Baldwin, Phillips, and Telervich (BPT) Diagnostic Diagrams and Maps}
\label{sec:3.3}

Both of AGN and starburst can drive galactic ionized winds, and consequently it is difficult to determine the powering source. 
The BPT diagnostic diagram \citep{1981PASP...93....5B} is often adopted to distinguish ionization and excitation mechanisms of ionized gas with similar kinematics. 
Generally it is assumed that the ionizing source of the gas also drive the outflow, so BPT diagram of ionized gas with similar kinematics is a popular diagnostic tool. 

In Figure~\ref{fig:kin_ha} and Figure~\ref{fig:kin_o3}, our results show that the kinematics of [O~{\sc{iii}}] and $\rm H\alpha$ are inconsistent for both the narrow and the broad components, which suggest that the BPT diagram with the two components cannot be used directly. Previous work by \cite{2021ApJ...906L...6R} derived the BPT diagram for the narrow component, and cautioned that the blueshifted component of [O~{\sc{iii}}] and $\rm H\beta$ have inconsistent kinematics for many spaxels ($\rm <2.\arcsec5$). 

\begin{figure}[htbp!]
	\includegraphics[width=\columnwidth]{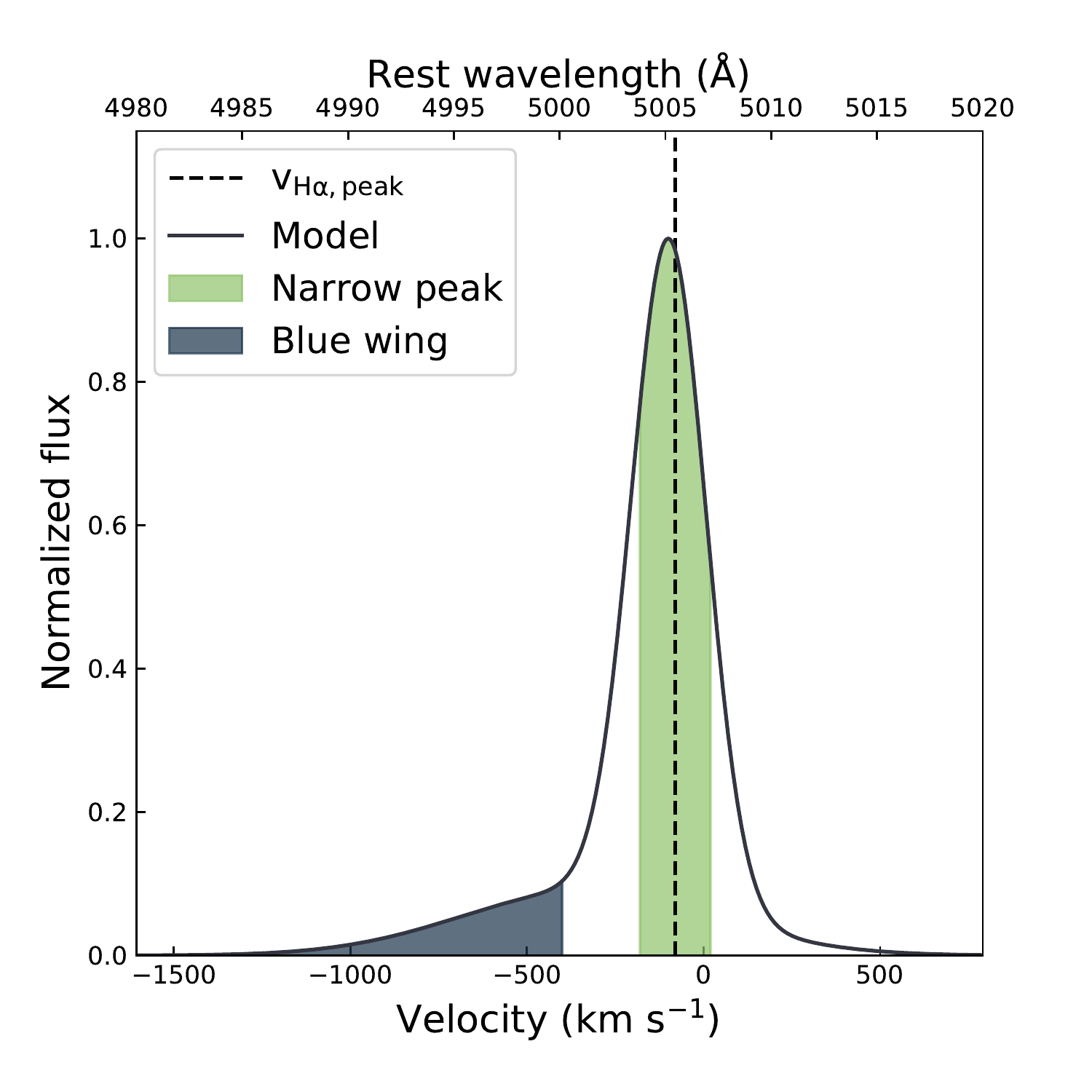}
	\caption{An illustration of the ``narrow peak" and the ``blue wing". 
	The ``narrow peak" is defined by the central portion with velocity $v_{\rm H\alpha,peak}-100< v < v_{\rm H\alpha,peak}+100\rm\ km\ s^{-1}$, which should be dominated by ionized gas from the disk. 
	The ``blue wing" is the portion with $v\ < -400\rm\ km\ s^{-1}$, which should be dominated by the emission from the outflow.
	The dashed line denote $v_{\rm H\alpha,peak}$ and the solid line is the synthetic model.
	}
	\label{fig:np&bw}
\end{figure}

To obtain the BPT diagnostic diagrams for the rotational disk and the outflow components, two intervals in the line profile are invoked, the ``narrow peak" and the ``blue wing".  As illustrated in Figure~\ref{fig:np&bw}, the ``narrow peak" is the portion of emission line with velocity $v_{\rm H\alpha,peak}-100< v < v_{\rm H\alpha,peak}+100\rm\ km\ s^{-1}$, which should be dominated by the rotational disk (see the $v_{\rm H\alpha,peak}$ map in Figure~\ref{fig:np_kin_ha}). 
The ``blue wing" is part of the line wing with $\rm v<-400\ km\ s^{-1}$, which is dominated by emission from the outflow component (see the $\rm \Delta v_{[O\, III]}-v_{H\alpha,peak}$ map in Figure~\ref{fig:np_kin_o3}). Such kinematically resolved BPT diagrams and maps has been demonstrated useful by previous works \citep[e.g.,][]{2019A&A...622A.146M}.

\begin{figure*}[htbp!]
	\gridline{\fig{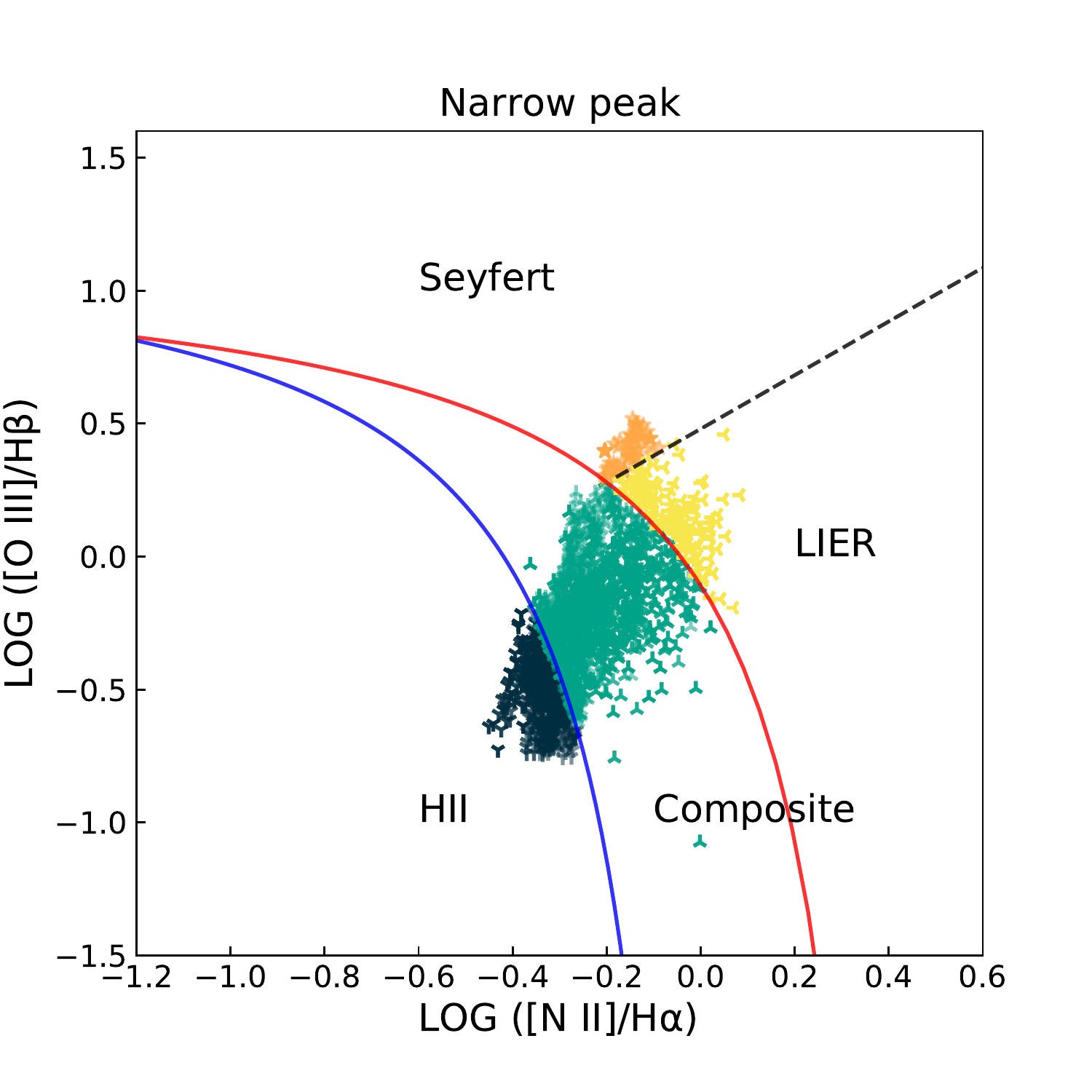}{0.45\textwidth}{(a)}
          \fig{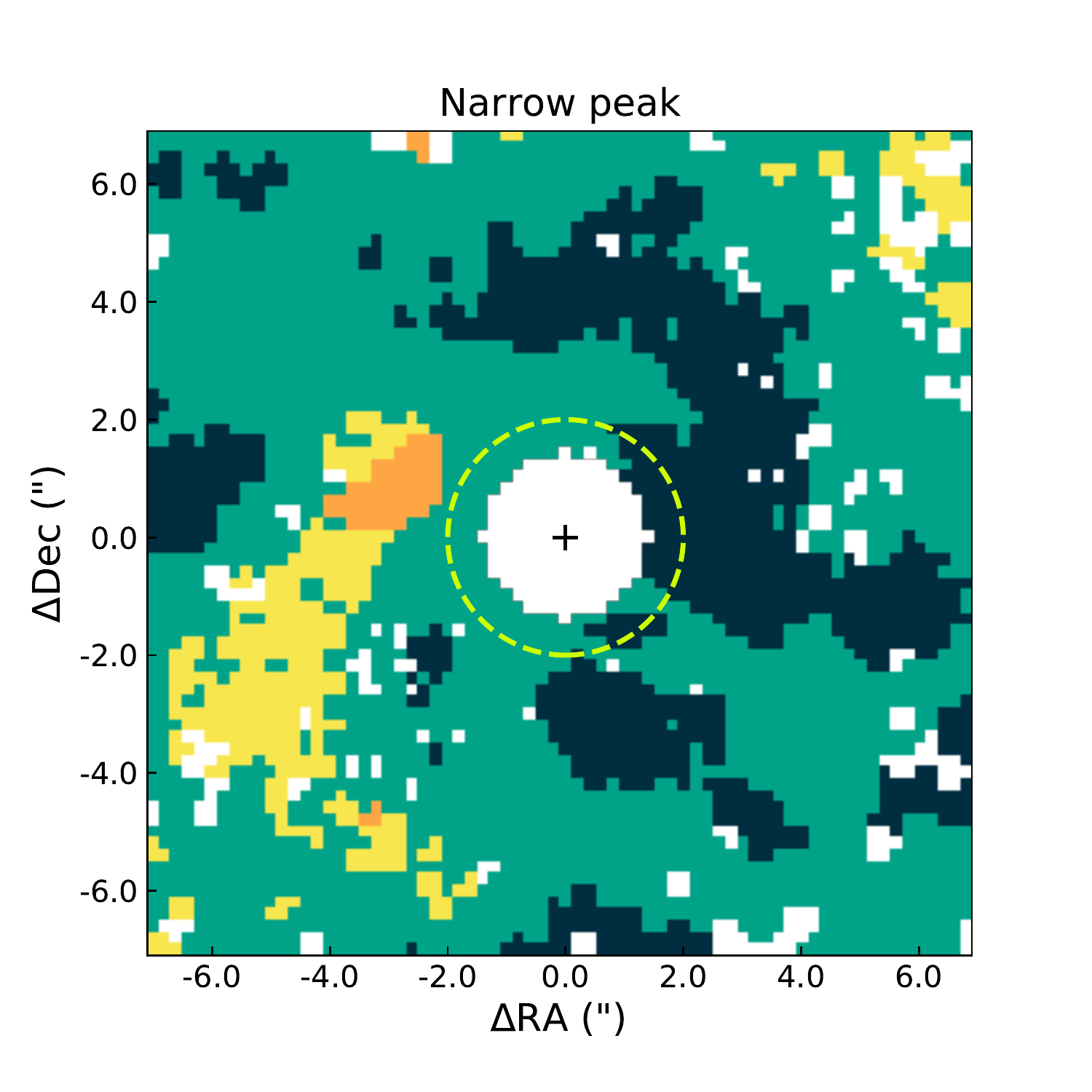}{0.45\textwidth}{(b)}}
          
    \gridline{\fig{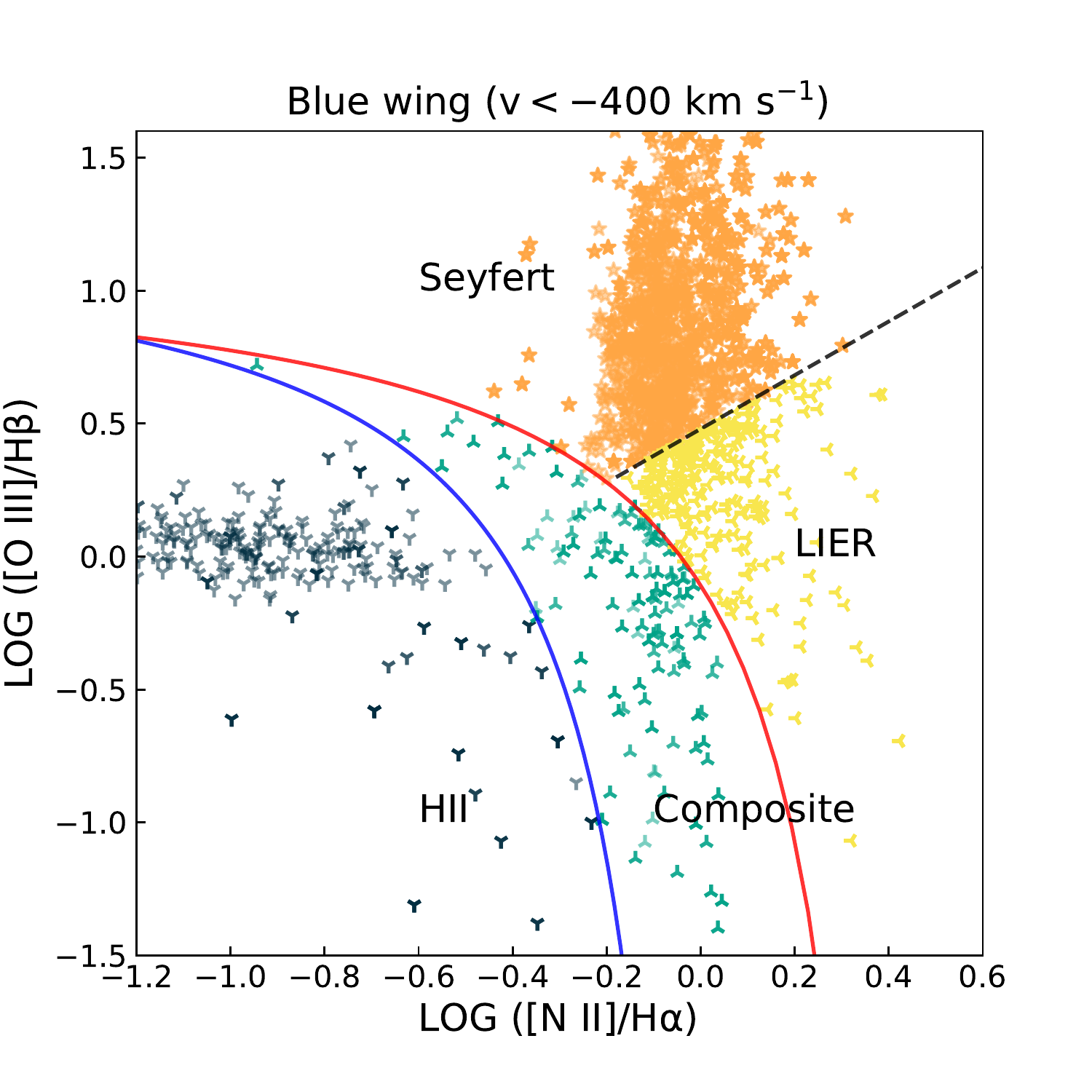}{0.45\textwidth}{(c)}
          \fig{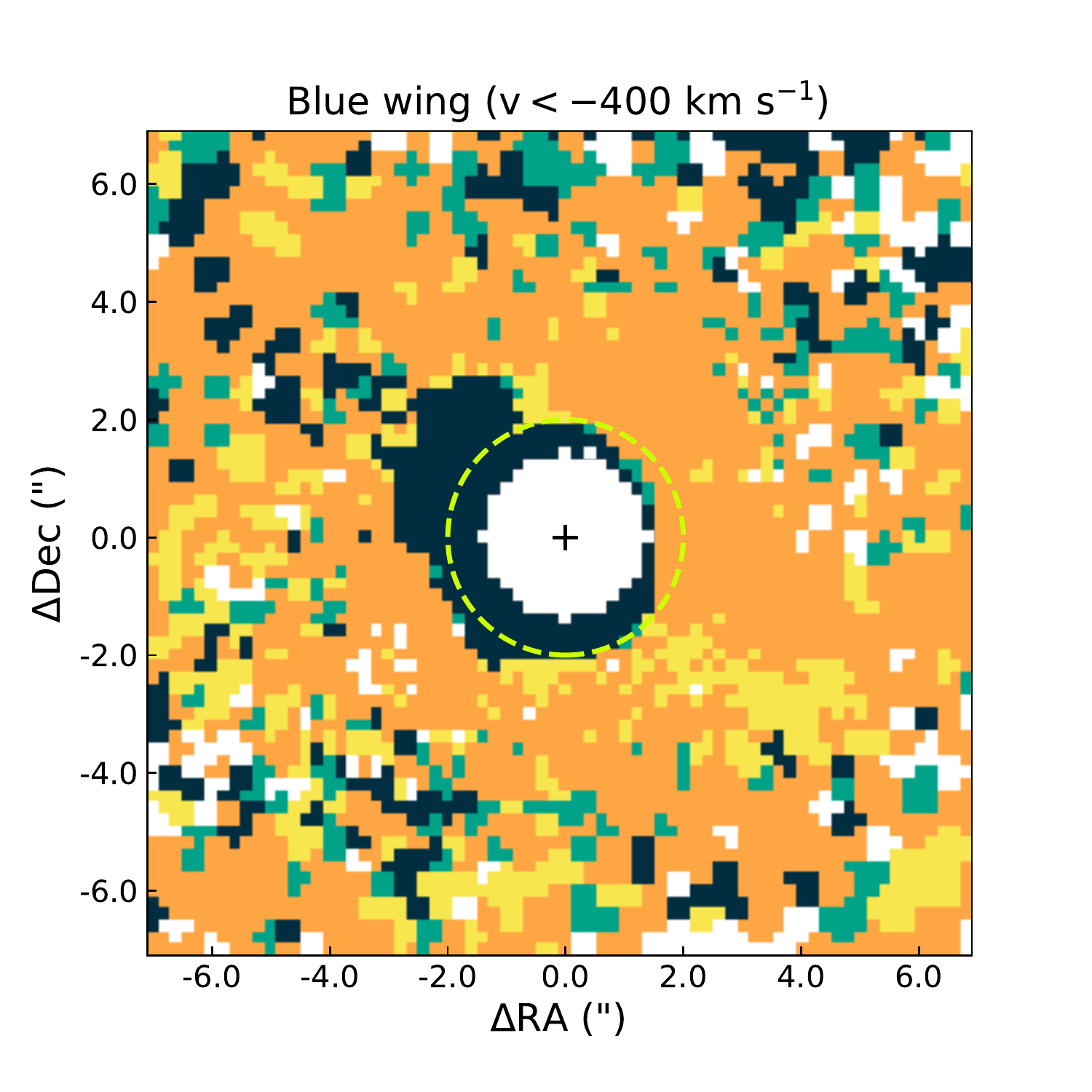}{0.45\textwidth}{(d)}}
    \caption{The BPT diagram and kinematically resolved map of the ``narrow peak" emission, shown in a and b, respectively. The same maps are shown for the ``blue wing" emission (c and d). The blue solid line \citep{2003MNRAS.346.1055K} separates the starforming and composite zones, the red solid line \citep{2001ApJ...556..121K} separates the composite and AGN/LI(N)ER, and the dashed line \citep{2010MNRAS.403.1036C} separates the AGN and LI(N)ER. 
    Nuclear regions ($<1.\arcsec4$) in these two maps are excluded due to the significant influence from BLR.
    North is to the top and east is to the left.
    }
    \label{fig:bpt}
\end{figure*}

Figure~\ref{fig:bpt}a and~\ref{fig:bpt}b show the BPT diagnostic diagram and the corresponding ionization map of the ``narrow peak" emission, respectively. Due to the exclusion of the nuclear region ($<1.\arcsec4$), the low ionization emission-line region (LIER) is used instead of the low ionization nuclear emission-line region (LINER) in this work. 
The ``narrow peak" BPT diagram (Figure~\ref{fig:bpt}a) shows a concentrated distribution in the composite region extending to the Seyfert/LIER and the star-forming region, which is expected given the coexistence of starburst and AGN in NGC 7469. A patchy spiral arm structure classified as H~{\sc{ii}} (star formation) ionization can be found in the ``narrow peak" BPT map, which is consistent with the spiral arm structure in the flux map of $\rm H\alpha$ (Figure~\ref{fig:flux_ha}). An intriguing arc structure mainly classified as LIER is also apparent in the southeast quadrant of the ``narrow peak" BPT map, $\rm \sim 2\arcsec$--$6\arcsec$ away from the nucleus.
A similar arc structure classified as LIER is also present in the ``blue wing" BPT map, corresponding to the same area but slightly further away from nucleus. 
According to previous work, such extended LIER regions could be associated with shocks produced by a jet or an outflow \citep[e.g.][]{2014MNRAS.444.3894H}. Some studies also suggested that LI(N)ER can usually be found along the wall of conical outflow in AGN \citep[e.g.][]{2016ApJ...829...46M,2019A&A...622A.146M}. The emission from the outflow can also contribute to the ``narrow peak" due to low line-of-sight velocity clouds in the outflow. We speculate that the arc structures in the ``narrow peak" and ``blue wing" BPT map could be associated with the edge of the kpc-scale ionization cone, which are affected by shocks produced by the outflow. Based on the morphology, we can estimate the projected outflow size of $\sim 2 \rm\ kpc$. 

Figures~\ref{fig:bpt}c and~\ref{fig:bpt}d show the BPT diagnostic diagram and map of the ``blue wing" emission. The ``blue wing" BPT diagram shows two domains of significantly clustered spaxels, one is H~{\sc{ii}} region and the other is the Seyfert/LIER region.
Figures~\ref{fig:bpt}d shows that most spaxels beyond the starburst ring are classified as AGN ionized region, which suggests that the outflow beyond the starburst ring are mainly powered by the central AGN. The regions around the starburst ring and extending to the northeast are dominated by H~{\sc{ii}} ionization, which indicates the outflow in these regions are likely driven by starburst.  This is consistent with the strong star forming activities in the ring and the findings of \cite{2021ApJ...906L...6R}.

\subsection{Mass and Energy Outflow Rates}
\label{sec:3.4}

\begin{deluxetable*}{ccccccc}
\tablenum{1}
\tablecaption{Properties of ionized outflowing gas\label{tab:rate}}
\tablewidth{0pt}
\tablehead{
\colhead{} & \colhead{$L_{\rm [O\, III]}$} & \colhead{$L_{\rm H\beta}$} & \colhead{$\dot{M}_{\rm out}$} & \colhead{$\dot{E}_{\rm out}$} & \colhead{$\dot{P}_{\rm out}$} & \colhead{Fraction}  \\
\colhead{Source} & \colhead{$\times10^{41}\rm\ erg\ s^{-1}$} & \colhead{$\times10^{41}\rm\ erg\ s^{-1}$} &
\colhead{$\rm M_{\astrosun}\ yr^{-1}$} & \colhead{$\times10^{41}\rm\ erg\ s^{-1}$} & \colhead{$\times10^{34}\rm\ g\ cm\ s^{-2}$} & \colhead{$\%$}  
}
\decimalcolnumbers
\startdata
Starburst & 4.6 & 4.8 & 6.8 & 5.5 & 2.2 & $63.3$ \\
AGN & 3.6 & 2.8 & 3.9 & 3.2 & 1.3 & $36.7$ \\
Total & 8.2 & 7.6 & 10.7 & 8.7 & 3.5 & - \\
\enddata
\tablecomments{(1) Ionization source of the outflowing gas. (2) Luminosity of [O\, {\sc{iii}}] associated with the outflow. (3) Luminosity of $\rm H\beta$ associated with the outflow. (4) Mass outflow rate with unit of $\rm M_{\astrosun}\ yr^{-1}$. (5) Energy outflow rate with unit of $\times10^{41}\rm\ erg\ s^{-1}$. (6) Momentum outflow rate with unit of $\times10^{34}\rm\ g\ cm\ s^{-2}$. (7) Fractions of outflow rates of ionized gas powered by AGN and starburst to total outflow rate.}
\end{deluxetable*}

The ionized gas outflow mass can be estimated following \cite{2006agna.book.....O}, which assumes the ``case B" recombination with electron temperature $\rm T_{e}=10^{4}K$:
\begin{equation}
	M_{\rm gas}=6.78\times10^{8} (\dfrac{L_{\rm H\beta}}{10^{43}\rm\ erg\ s^{-1}})(\dfrac{n_{e}}{100\rm\ cm^{-3}})^{-1}\ M_{\astrosun}, \label{eq:1}
\end{equation}
where $L_{\rm H\beta}$ is the luminosity of $\rm H\beta$, $n_{\rm e}$ is the electron density which can be calculated from the ratio [S~{\sc{ii}}]$\lambda 6716/$[S~{\sc{ii}}]$\lambda 6731$. 
Assuming the electron temperature $T_{e} = 10^{4}$ K, the mean electron density across the FoV is estimated to be $n_{\rm e}\sim380\rm\,cm^{-3}$.
The mass, energy and momentum outflow rate are calculated following:
\begin{align}
    \dot{M}_{\rm out} &= 3M_{\rm gas}\dfrac{v_{\rm out}}{R_{\rm out}}, \label{eq:2}\\
    \dot{E}_{\rm out} &= \dfrac{1}{2}\dot{M}_{\rm out}v_{\rm out}^{2}, \label{eq:3}\\
    \dot{P}_{\rm out} &= \dot{M}_{\rm out}v_{\rm out}, \label{eq:4}
\end{align}
where $v_{\rm out}$ denotes the averaged velocity of the outflow, and $R_{\rm out}$ is the outflow size.

According to the kinematics maps of $\rm H\alpha$, the non-rotational component cannot fully represent the outflow. Consequently, the broad component of $\rm H\beta$ cannot be directly used to estimate the total mass of the outflowing ionized gas. 
The $\rm H\beta$ flux associated with the outflow component ($F_{\rm H\beta,OF}$) can be estimated following:
\begin{align}\label{eq:5}
	\left\{
	    \begin{array}{ll}
	F_{\rm [O\, III],R} &= F_{\rm H\beta,R} \times \dfrac{F_{\rm [O\, III],R}}{F_{\rm H\beta,R}} 
	\\ 
	F_{\rm [O\, III],OF} &= F_{\rm [O\, III],tot} - F_{\rm [O\, III],R} 
	\\ 
	F_{\rm H\beta,OF} &= F_{\rm [O\, III],OF} \times \dfrac{F_{\rm H\beta,OF}}{F_{\rm [O\, III],OF}}\ ,
		\end{array}
	\right.
\end{align}
where F denotes the flux, the subscripts [O~{\sc{iii}}] and $\rm H\beta$ represent different emission lines, and the subscripts R, OF and tot denote rotational component, outflow component and total flux, respectively.
The total luminosity of $\rm H\beta$ and [O~{\sc{iii}}] associated with the outflow component can be estimated to be $L_{\rm H\beta,OF} = 7.6\times10^{41}\rm\ erg\ s^{-1}$ and $L_{\rm [O\, III],OF} = 8.2\times10^{41}\rm\ erg\ s^{-1}$.

In NGC 7469, both the nuclear starburst ring and the AGN are the powering source responsible for the total mass and energy outflow rates. 
To estimate the outflow rates, we assume that the outflow in spaxels classified as H~{\sc{ii}} in the BPT diagram are driven by the starburst and the outflow in other spaxels are driven by AGN (see Figure~\ref{fig:bpt}d). 
\citet{2021ApJ...906L...6R} suggest that the most of the ionized outflow in the circumnuclear region ($r<2.\arcsec5$) are possibly driven by starburst.
We use the mean value of [O~{\sc{iii}}]/$\rm H\beta$ of those pixels classified as the H~{\sc{ii}} in the BPT diagram as the [O~{\sc{iii}}]/$\rm H\beta$ ratio of the masked nuclear region ($r<1.\arcsec4$).

The luminosity of $\rm H\beta$ associated with the starburst wind is estimated at $L_{\rm H\beta,SB} = 4.8\times10^{41}\rm\ erg\ s^{-1}$ according to the equation~\ref{eq:5} and an assumption of the biconical outflow geometry.
The luminosity of $\rm H\beta$ produced by the AGN outflow is $L_{\rm H\beta,AGN} = L_{\rm H\beta,tot}-L_{\rm H\beta,SB} = 2.8\times10^{41}\rm\ erg\ s^{-1}$. 
Note that $L_{\rm H\beta,SB}$ can be overestimated under our assumptions due to ignoring the AGN contribution in the masked region.
For simplicity, the sizes of the starburst wind and the AGN outflow are both assumed to be $R_{\rm out}\sim 2\rm\ kpc$.
The average $v_{10} \sim -510\rm\ km\ s^{-1}$ and electron density $n_{\rm e} \sim 380 \rm\ cm^{-3}$ across the FoV are used to estimate the outflow rates.
Based on Equations (\ref{eq:1}）--（\ref{eq:4}), the mass, energy and momentum outflow rate of the starburst wind and the AGN outflow are estimated and summarized in Table~\ref{tab:rate}.  Outflow rates from starburst are larger than the ones from AGN, implying that starburst-driven outflows are more powerful than AGN-driven outflows in NGC 7469. Nevertheless, the AGN-driven outflows are also non-negligible, carrying $\rm >30\%$ of the total outflow mass and energy rates.

It is worth noting that in recent years, many works have pointed out the estimated mass and energy outflow rates may vary by one to two orders of magnitude when using different assumptions and derived quantities such as ionized gas mass, velocities, and densities \citep[e.g.][]{2021MNRAS.504.3890D}.  Our work also cannot avoid these uncertainties, and the coexisting starburst and AGN outflows further complicate the measurement. However, the first order estimated outflow rates and the relative contribution are still useful for our understanding of impact of the starburst wind and the AGN outflow.

\section{Discussion}
\label{sec:4}

\subsection{The AGN-driven outflow}
\label{sec:4.1}

In NGC 7469, a biconical outflow across the east-west direction traced by [Si VI]$\lambda$ 1.96$\mu$m emission in the near-infrared was previously discovered by \cite{2011ApJ...739...69M}. \cite{2021ApJ...906L...6R} recently reported discovery of a patch of fast outflow in [O~{\sc{iii}}] in the circumnuclear region possibly driven by AGN. The location is to the northwest of the nucleus, which appears to be associated with the blueshifted edge of the outflow in \cite{2011ApJ...739...69M}.  We have found sub-galactic scale AGN-driven outflow extending to at least $\sim 2 \rm\ kpc$, which resembles a blueshifted tilted cone outflowing towards us. 
The direction of the larger scale AGN-driven outflow does not fully correspond to the circumnuclear outflow in the inner few arcseconds.

One possible explanation of the misalignment is the projection effect and the contamination of starburst-driven outflows.
NGC 7469 is a nearly face-on Seyfert 1 galaxy with an inclination angle of $\sim30^{\circ}$, which implies that the starburst wind is heavily affected by the projection along the line-of-sight. 
The inclination angle of the AGN outflow has been modeled by \cite{2011ApJ...739...69M} with $\sim 57^{\circ}$ (the blueshifted cone points to the west) using [Si VI] emission on the sub-arcsecond scale, which also suggests that the projection effect can have influence on the AGN outflow.
The detailed geometry of outflow requires more spatially resolved data with higher S/N, which is beyond the capability of current data.
Besides the projection effect, the emission of the starburst wind could overwhelm the emission from the AGN-driven outflow around the starburst ring. 
In the ``blue wing" BPT map (Figure~\ref{fig:bpt}d), the H~{\sc{ii}} ionization is more extended in the eastern region (A5, A6, and A7) than in the west (A1, A2, and A3) around the starburst ring, which indicates that emission from the AGN outflow in the eastern region is weaker than western region.
The asymmetric features of optical line emissions (Figure~\ref{fig:asymmetry}) are consistent with this scenario.
This may well explain why \cite{2021ApJ...906L...6R} only found outflows from a small region in the west region inside the starburst ring, which was attributed to be driven by AGN. 
Moreover, \cite{2021ApJ...906L...6R} distinguish the AGN outflow only based on the kinematics and the BPT diagram of narrow component emission, whilst our approach is complementary to identify outflow with broad wings in emission lines.  
In reality, it is more likely that the AGN outflow and the starburst wind both contribute to the kpc scale outflow in NGC 7469.  
The outflowing gas driven by starburst could also become photoionized by the AGN once entering the AGN ionization cone, and the line ratios may be altered to values corresponding to AGN ionization (see Section~\ref{sec:4.2}).

The diffuse soft X-ray emission in NGC 7469 has been found by \cite{2018A&A...615A..72M} using the longest {\em Chandra} observation (see Section~\ref{sec:xray}). 
They suggested that the diffuse soft X-ray emission is coronal emission from nuclear starburst ring.  
Our results show that the ionized outflows driven by AGN reach to $\rm \sim 2$ kpc, which is fully in agreement with the size of the diffuse soft X-ray emission (see Figure~\ref{fig:flux_o3} and Figure~\ref{fig:bpt}). 
The asymmetry of soft X-ray is also consistent with [O III] emission generally (Figure~\ref{fig:asymmetry}).
We have shown in Figure~\ref{fig:bpt} that the starburst wind is mainly concentrated in the circumnuclear regions in and around the starburst ring. 
Therefore, it is plausible that the AGN-driven outflow may contribute to a part of this diffuse soft X-ray emission especially on kpc scale, besides the main emission from the starburst ring.  

\begin{figure*}[htbp!]
	\includegraphics[width=\textwidth]{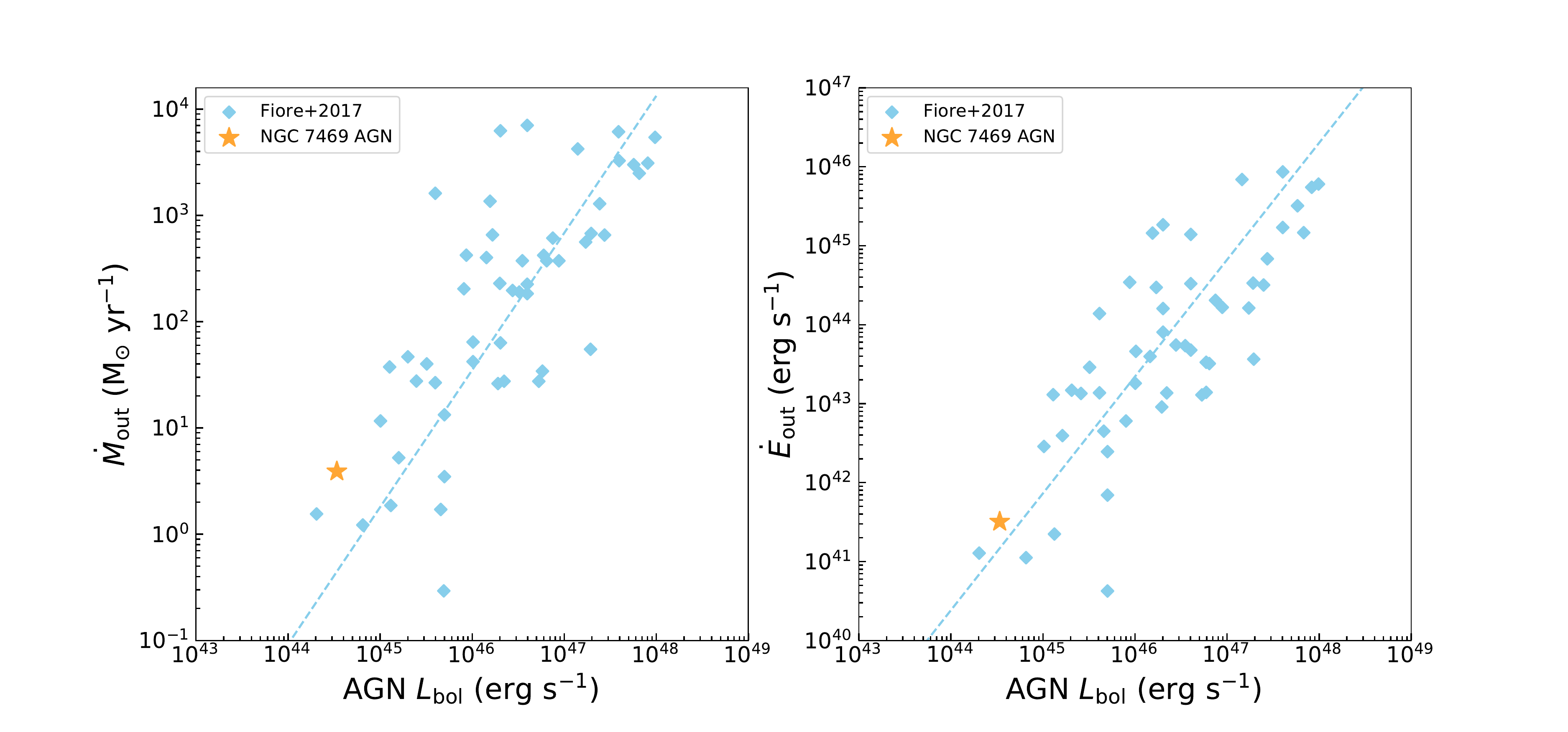}
    \caption{Mass (left panel) and energy (right panel) outflow rate as a function of the AGN bolometric luminosity. The diamonds denote the ionized outflow measurements from \citet{2017A&A...601A.143F} and the dashed line is their best-fit correlation. The stars in two panels are our estimate of AGN ionized gas outflow rates.}
    \label{fig:out func}
\end{figure*}

Figure~\ref{fig:out func} shows the outflow rates as a function of the AGN bolometric luminosity following \citet{2017A&A...601A.143F}. 
Our measurement of the NGC 7469 AGN outflow are cosnsistent with the best-fit correlation.
This implies the properties of NGC 7469 outflow is rather typical of the sources in \citet{2017A&A...601A.143F}.

Theoretically the AGN outflow can be energy-driven or momentum-driven according to the different cooling efficiency \citep[e.g.][]{2012ApJ...745L..34Z}. 
Considering the bolometric luminosity of AGN $L_{bol}=10^{44.53}\rm\, erg\, s^{-1}$ \citep[][]{2012A&A...542A..83P}, the momentum of AGN radiation is $\dot{P}_{\rm AGN} = L_{\rm AGN,bol}/c = 1.1 \times 10^{34} \rm \,g\,cm\,s^{-2}$.  
As shown in Table~\ref{tab:rate}, the momentum flux of the kpc-scale AGN outflow is $\dot{P}_{\rm AGN,out} = 1.3 \times 10^{34} {\rm \,g\,cm\,s^{-2}} \sim L_{\rm AGN,bol}/c$, which is consistent with the momentum-driven scenery.
The kinetic coupling efficiency of AGN-driven outflows can be estimated as $\varepsilon \sim \dot{E}_{\rm out}/L_{\rm bol}\sim0.09\%$ which is lower than the threshold of $0.5\%$ suggested by the ``two-stage" feedback model by \cite{2010MNRAS.401....7H}.
The low value of the kinetic coupling efficiency might suggest that the kpc-scale AGN outflow does not significantly affect the growth of the host galaxy.
Note that the kinetic coupling efficiency estimated here is only based on ionized gas outflow driven by AGN, other phases of outflows and starburst winds are not included.   
Whereas cold gas outflow in NGC~7469 can be important, previous works \citep[e.g.][]{2015ApJ...811...39I} have shown no significant evidence of molecular or neutral outflows has been found.

\subsection{The starburst wind}
\label{sec:4.2}

The energy outflow rate of the starburst driven wind can be expected by the equation as follow \citep[][]{2017A&A...601A.143F}:
\begin{equation}
	\dot{E}_{\rm SN} = \eta \times {\rm SFR} \times \psi \times \xi_{\rm SN} \times L_{\rm SN},
	\label{eq:6}
\end{equation}
where $\psi$ denotes the fraction of massive stars under certain initial mass function (IMF), $\xi_{SN}$ is the number of supernova (SN) per solar mass, $L_{\rm SN}$ is the total luminosity for each SN, and $\eta$ denotes the efficiency from total luminosity transform to kinetic power of ISM.
Assuming the Salpeter IMF, $\xi_{SN}=0.0066 \rm\,SNe/M_{\astrosun}$, $L_{\rm SN}=10^{51}\rm\, erg\,s^{-1}$, and $\eta=0.1$, $\dot{E}_{\rm SN}$ is expected at $\sim 10^{42}\rm\, erg\,s^{-1}$ adopting $\rm SFR \sim 48\,M_{\astrosun}\,yr^{-1}$ \citep[][]{2011A&A...535A..93P}. 
This value is broadly consistent with $\dot{E}_{\rm out,SB}=0.55\times 10^{42}\rm\, erg\,s^{-1}$ estimated by us, which indicates the SN process can power the starburst wind in NGC 7469.

Combining the [O~{\sc{iii}}] kinematics maps (Figure \ref{fig:np_kin_o3}) and the ``blue wing" BPT map (Figure \ref{fig:bpt}), the kinematics of the starburst wind in the east region around the starburst ring shows lower velocity than the AGN-driven outflows in the west region, which is consistent with the findings of \cite{2021ApJ...906L...6R}. 
However, considering the outflow rates (see Table \ref{tab:rate} and Section \ref{sec:3.4}), the starburst wind is more powerful than the AGN-driven outflow due to the higher outflowing mass in the starburst wind.  These results indicate that the starburst wind could show slower kinematics but energetically more powerful than the AGN-driven outflow in NGC 7469.

According to the ``blue wing" BPT map, impact of the starburst wind is concentrated around the starburst ring, and outflows on the larger scale ($\rm \sim 2kpc$) are mainly driven by AGN.  
This result may suggest that the starburst mainly drives a wind on small scale (100 pc) in NGC 7469, whilst on the larger scale (kpc) the AGN becomes the main powering source driving the outflow. 
Such a dynamic scenario suggests that the ionized outflow in NGC 7469 should be a joint contribution from stellar and AGN feedback. 
Ionized gas clouds in the starburst ring are constantly blown out of the disk by starbursts. When these clouds enter the AGN ionization cone and are far away from the starburst ring, they will be ionized and become photoionized by the AGN.

In many nearby Seyfert galaxies, a starburst nuclear ring and a galactic outflow can be found similar to NGC 7469.
For example, a star formation ring and a kpc-scale biconical ionized outflow can be found in Seyfert 1.8 galaxy NGC 1365 \citep[see][and references therein]{2009ApJ...694..718W,2018A&A...619A..74V}.
Because the biconical outflow is mainly classified as AGN photoionization in the BPT diagram and its mass outflow rate is similar to that of nuclear AGN wind detected by X-ray, \citet{2018A&A...619A..74V} attributed the driving source of the outflow to AGN. Using the kinematically resolved BPT diagram and map, \citet{2019A&A...622A.146M} further confirmed that the ionization source of the kpc-scale outflow in NGC 1365 is dominated by AGN, and the line ratios of the outflow are dominated by SF ionization only in the starburst ring \citep[see Fig.D.1 in][]{2019A&A...622A.146M}. 
Compared to NGC 1365 \citep[$\rm SFR \sim 5.6\,M_{\astrosun}\,yr^{-1}$,][]{2012MNRAS.425..311A}, the starburst wind of NGC 7469 is more energetic, which might be attributed to the higher star formation rate of NGC 7469 ($\rm SFR \sim 48\,M_{\astrosun}\,yr^{-1}$).  
Both galaxies contain circumnuclear starburst rings and AGNs in the nuclear region, their outflows can be driven by different sources, i.e. AGN and starburst.

In most previous studies, outflows in galaxies are often assumed to be driven by a single powering source (AGN or starburst) for simplicity, which seems appropriate for galaxies like NGC 1365 but not the ones like NGC 7469.  Our results suggest that the starburst wind and the AGN outflow can both be important in some cases in terms of energetics.  In the local universe, many (U)LIRGs containing both starburst and AGN activities are rather similar to NGC 7469, which can possess both powerful starburst winds and the AGN outflow.  In the case of high redshift quasars, it is generally expected that nuclear starburst and AGN activities can both be more powerful than local Seyfert galaxies \citep[e.g.][]{2012ARA&A..50..455F,2020ARA&A..58..661F}.  NGC 7469 is an excellent example for further studying the feedback from both SF and AGN processes.

\subsection{Putative evidence of inflows}
\label{sec:4.3}
As shown in Figure~\ref{fig:np_kin_ha}, $\rm \Delta v_{H\alpha} - v_{H\alpha,peak}$ and $\rm W80_{H\alpha}$ maps of $\rm H\alpha$ present a redshifted, possibly inflow-like feature in the northeast and east regions around the starburst ring.  Theoretical work and many observations suggested that a bar-like structure in galaxies can dynamically drive molecular gas inflows and enhance star formation in the central region \citep[e.g.][]{2011MNRAS.416.2182E,2012MNRAS.423.3486W}. 
A bar-like or spiral structure in molecular gas along northeast to southwest has been detected by \cite{2004ApJ...602..148D}, which is partially coincident with the circumnuclear inflow-like structure connecting the CND and the starburst ring . However, no stellar bar is found in the infrared \citep[][]{1995ApJ...444..129G,2000AJ....119..991S} or radio \citep[][]{1991ApJ...378...65C,2010MNRAS.401.2599O}.
\cite{2004ApJ...602..148D} also reported that no molecular gas inflow or other kinematics characteristics expected for a bar have been observed along the bar-like structure. \cite{2015ApJ...811...39I} further suggest that this feature is possibly not a bar but several spiral structures blending for the lack of spatial resolution.  Nevertheless, \cite{2015ApJ...811...39I} noted that the existence of a faint stellar bar cannot be ruled out.  In addition to the bar, the tidal interaction due to galaxies merging can also drive gas inflows and starburst activities in the central region of galaxies. In our case, NGC 7469 and IC 5283 are undergoing the merger process, and the nuclear starburst ring in NGC 7469 is believed to be produced by the tidal interaction with IC 5283 \citep[][]{1995ApJ...444..129G}. We suggest the inflow-like features in the $\rm H\alpha$ non-parametric kinematics maps might not be driven by the bar, but instead the tidal interaction between NGC 7469 and IC 5283. 

There exists another inflow-like feature in the southeast region of the FoV, approximately $4\arcsec$ away from the nucleus (Figure~\ref{fig:np_kin_ha}).  In the ``narrow peak" BPT map, this region is dominated by LIER and composite emission, while star forming, LIER, and AGN all exist in this region in the ``blue wing" BPT map. It is likely that a stream of cold gas is falling into the disk, and spatially superimposed by outflows from the AGN. However, the quality of current data is insufficent to further verify this scenario.

\section{Conclusions}
\label{sec:5}

In this work, we study the spatial resolved ionized gas nature of NGC 7469 extending the FoV to $\rm 14^{\prime \prime}\times14^{\prime \prime}$ using the archival VLT/MUSE data. 
Two Gaussian fitting and non-parametric approaches are used to analyze the datacube. 
Zeroth-order image of soft X-ray (0.2--1.0 keV) from all available {\em Chandra} archive data is merged. Taking into account of the nuclear PSF model, the morphology of extended X-ray emission is obtained (Figure~\ref{fig:xray}).   
To identify the origins of ionized gas outflows, ``narrow peak" ($v_{\rm H\alpha,peak}-100< v < v_{\rm H\alpha,peak}+100\ \rm km\ s^{-1}$) and ``blue wing" ($v\ < -400\rm\ km\ s^{-1}$) are defined to represent rotational disk and outflows components when using BPT diagnostic diagrams and maps (Figure~\ref{fig:bpt}).  The main results and conclusions are summarized as follows:

\begin{enumerate}
    \item The kinematics of [O~{\sc{iii}}] and Balmer emission lines show different patterns, which is consistent with previous studies. 
    The former is dominated by the outflowing gas and the later shows rotational disk superposed with complex non-rotational components.
    
    \item An outflow extending to $\rm \sim2$ kpc in projection traced by [O~{\sc{iii}}] is found in NGC 7469 from the MUSE data, which is mainly ionized by AGN according to the ``blue wing" BPT diagram and map (Figure~\ref{fig:bpt}).
    The geometry of the kpc-scale AGN outflow does not fully match the smaller scale AGN outflow detected in the nuclear region \citep[][]{2011ApJ...739...69M,2021ApJ...906L...6R}, which might be due to projection effect and contamination of the emission from the starburst ring.
    
    \item The morphology of the kpc-scale outflow is consistent with that of the extended soft X-ray (Figure~\ref{fig:flux_o3}), which implies that the AGN outflow may partly contribute to the extended soft X-ray emission. However, most this soft X-ray emission should be associated with star formation due to the star forming ring and the intense starburst wind.
    
    \item The mass and energy outflow rates of the AGN outflow are estimated to be $\dot{M}_{\rm out,AGN} \sim 3.9 \rm\ M_{\astrosun}\ yr^{-1}$ and $\dot{E}_{\rm out,AGN} \sim 3.2 \times 10^{41} \rm\ erg\ s^{-1}$ (see Section~\ref{sec:3.4}).
    Though the outflow rates of the AGN outflow are less than that of the starburst wind, the AGN outflow are still important energetically by carrying $\rm >30\%$ of the total outflow rates. 
   The kinetic coupling efficiency of the AGN-driven outflow of $\varepsilon=\dot{E}_{\rm out}/L_{\rm bol}\sim0.1\%$ is lower than the threshold of $0.5\%$ suggested by the feedback model in \cite{2010MNRAS.401....7H}, implying a marginal impact by the detected kpc-scale outflow to the growth of the host galaxy.
    
    \item The starburst wind around the nuclear ring is also confirmed, consistent with the findings in \cite{2021ApJ...906L...6R}. 
    The mass and energy outflow rates of the starburst wind are estimated to be $\dot{M}_{\rm out,SB} \sim 6.8 \rm\ M_{\astrosun}\ yr^{-1}$ and $\dot{E}_{\rm out,SB} \sim 5.5 \times 10^{41}\rm\ erg\ s^{-1}$.  
    In NGC 7469, starburst seems to be driving outflow on a smaller scale than AGN (Figure~\ref{fig:bpt}), yet the starburst wind is more powerful than the AGN outflow in terms of higher outflow rate.
    
    \item Inflow-like patterns are found around the starburst ring extending to the northeast and in the southeast region, about a kpc distance away from the nucleus in projection. If these features are streams of inflow, the tidal interaction between NGC 7469 and IC 5283 might be the driving source.
\end{enumerate}

Last but not least, many local (U)LIRGs and high redshift quasars can possess both powerful starburst winds and AGN outflows like NGC 7469.  Hence, NGC 7469 is an excellent local example for studying the joint processes of feedback from SF and AGN.

\begin{acknowledgments}
We thank the anonymous referee for the constructive comments that significantly improved the clarity of our work. We thank Dr. Cheng Li, Yanmei Chen, and Zhenzhen Li for helpful suggestions on data analysis and discussion. J.W. acknowledges the NSFC grants U1831205 and 12033004, and the science research grants from CMS-CSST-2021-A06 and CMS-CSST-2021-B02. This research has made use of the services of the ESO Science Archive Facility. Based on observations collected at the European Southern Observatory under ESO programme 60.A-9339. This research has made use of data obtained from the Chandra Data Archive and the Chandra Source Catalog, and software provided by the Chandra X-ray Center (CXC) in the application packages CIAO and Sherpa. We acknowledge the allocated P200/CWI observation of NGC 7469 from the Telescope Access Program (TAP) for the PhD thesis of Xiaoyu Xu and the kind support by the staff at the Palomar Observatory.
\end{acknowledgments}

\vspace{5mm}
\facilities{VLT (MUSE), CXO (HETGS)}

\software{astropy \citep{2013A&A...558A..33A,2018AJ....156..123A}, PPXF \citep{2004PASP..116..138C}, MPFIT \citep{2009ASPC..411..251M}, LZIFU \citep{2016Ap&SS.361..280H}, CIAO \citep{2006SPIE.6270E..1VF}, DS9 \citep{2003ASPC..295..489J}, ChaRT \citep{2003ASPC..295..477C}, MARX \citep{2012SPIE.8443E..1AD}, Sherpa \citep{2001SPIE.4477...76F}
          }

\bibliography{7469}{}
\bibliographystyle{aasjournal}



\end{CJK*}
\end{document}